\documentclass[letterpaper,11pt]{article}
\pdfoutput=1
\usepackage{jheppub}
\usepackage{amsmath}
\usepackage{amssymb}
\usepackage{epsfig}
\usepackage{graphicx}
\usepackage{hyperref}
\usepackage{tikz}
\usetikzlibrary{decorations.pathmorphing}
\usepackage{pgfplots}
\pgfplotsset{width=6cm}%, compact=1.12}
\usepackage{graphicx}

\newcommand{\lie}{\pounds}

\newcommand{\M}{\mathcal{H}}
\newcommand{\MM}{\mathcal{M}}
\newcommand{\hp}{\mathbb{H}^{d-1}}
\newcommand{\ts}{\tilde{\sigma}}

\newcommand{\Z}{\mathbb{Z}}
\newcommand{\OO}{\mathcal{O}}
\newcommand{\tO}{\tilde{\mathcal{O}}}

\newcommand{\x}{\vec{x}}

\newcommand{\tr}{\text{tr} }

\renewcommand{\S}{\widetilde{S}}

\renewcommand{\r}{{\rho}_{{(n)}}}

\newcommand{\comment}[1]{}
\newcommand{\ket}[1]{\left|#1\right>}% | >
\newcommand{\bra}[1]{\left\langle#1\right|}% < |

\newcommand{\be}{\begin{equation}}% begin equation
\newcommand{\ee}{\end{equation}}% end equation

\title{The holographic dual of R\'enyi relative entropy}
\author[a]{Ning Bao,}
\author[b]{Mudassir Moosa,}
\author[b]{and Ibrahim Shehzad}

\affiliation[a]{Berkeley Center for Theoretical Physics, Berkeley, CA, 94720, USA}%\\
\affiliation[a]{Computational Science Initiative, Brookhaven National Lab, Upton, NY, 11973, USA}
\affiliation[b]{Department of Physics, Cornell University, Ithaca, NY, 14853, USA}

\emailAdd{ningbao75@gmail.com}
\emailAdd{mudassir.moosa@cornell.edu}
\emailAdd{is354@cornell.edu}

\abstract{The relative entropy is a measure of the distinguishability of two quantum states. A great deal of progress has been made in the study of the relative entropy between an excited state and the vacuum state of a conformal field theory (CFT) reduced to a spherical region. For example, when the excited state is a small perturbation of the vacuum state, the relative entropy is known to have a universal expression for \textit{all} CFT's \cite{Faulk-GR-entanglement}. Specifically, the perturbative relative entropy can be written as the symplectic flux of a certain scalar field in an \textit{auxiliary} AdS-Rindler spacetime \cite{Faulk-GR-entanglement}. Moreover, if the CFT has a semi-classical holographic dual, the relative entropy is known to be related to conserved charges in the bulk dual spacetime \cite{lashkari2016gravitational}. In this paper, we introduce a one-parameter generalization of the relative entropy which we call \textit{refined} R\'enyi relative entropy. We study this quantity in CFT's and find a one-parameter generalization of the aforementioned known results about the relative entropy. We also discuss a new family of positive energy theorems in asymptotically locally AdS spacetimes that arises from the holographic dual of the refined R\'enyi relative entropy. }

\begin{document}
\maketitle
%\tableofcontents((1+1))

\section{Introduction}

The introduction of quantum information theory into quantum gravity in general and the AdS/CFT correspondence in particular has led to a revolution in the latter. Specifically, formulating questions in gravity in terms of entropy measures such as the entanglement entropy and relative entropy has proven to be a very fruitful endeavor.
%\newline

Relative entropies are a way of quantifying the distinguishability of two quantum states, borne out of quantum information theory. The relative entropy between two states $\rho$ and $\sigma$ is defined as
\begin{align}
S_{\text{rel}}(\rho||\sigma) \, \equiv \, \tr( \rho\log\rho ) - \tr (\rho\log\sigma) \, . \label{eq-re-def-intro}
\end{align}
%In particular, the R\'enyi entropy between two states $\rho$ and $\sigma$ $S(\rho||\sigma)$ is given by
%\be
%S(\rho||\sigma)=\Tr \rho \Tr \log \rho - \Tr \rho \log \Tr \sigma.
%\ee
This quantity has proven to be very important in the context of quantum gravity, appearing in the proof of the Bekenstein bound \cite{Casini:2008cr,Longo:2018zib}, proofs of the generalized second law \cite{Wall:2010cj,wall2012proof}, proofs of the Bousso bound \cite{Bousso:2014sda,Bousso:2014uxa}, the proof of the average null energy condition \cite{faulkner2016modular}, study of the quantum null energy condition \cite{koeller2018local, leichenauer2018energy, Ceyhan:2018zfg}, study of constraints satisfied by the renormalization group trajectories in field theory \cite{add-ref-3}, the question of black hole microstate distinguishability \cite{BO}, the proof of entanglement wedge reconstruction \cite{JLMS, dong2016reconstruction, faulkner2017bulk, cotler2017entanglement}, the derivations of bulk constraints such as Einstein equations \cite{Lashkari:2013koa,Faulkner:2013ica,Banerjee:2014oaa,Swingle:2014uza,Banerjee:2014ozp}, and the recent holographic proofs of the positive energy theorems \cite{lin2014tomography,lashkari2015inviolable,Lashkari:2015hha,lashkari2016gravitational}. %\tcb{(citations.)}%In particular, relative entropy seems to be important to any question that relies on distinguishability between two different quantum states, where one of the quantum states considered in this context is often either the vacuum state or the state corresponding to some classical bulk geometry and the other state is some perturbation thereof. 
%\newline

%\tcb{Brief review of known results}\\
A significant effort has been made to study relative entropy in the context of conformal field theories (CFT's) on $R^{1,d-1}$, especially when the reference state is taken to be the vacuum state reduced to a spherical region. The relative entropy between an excited state ($\rho$) and the vacuum state $(\sigma)$ in this setup has been studied using field theory methods in \cite{Faulkner:2014jva,Lashkari:2014yva,Faulk-GR-entanglement} and using the AdS/CFT correspondence in \cite{Blanco:2013joa,Lashkari:2013koa,Faulkner:2013ica,Banerjee:2014oaa,Swingle:2014uza,Banerjee:2014ozp,lin2014tomography,lashkari2015inviolable,lashkari2016gravitational,Lashkari:2015hha}. In the case where the state $\rho$ in Eq.~\eqref{eq-re-def-intro} can be written as small perturbation around the vacuum state $\sigma$, the relative entropy (to lowest order in the perturbation parameter) can be written in terms of the CFT two point functions \cite{Faulkner:2014jva,Faulk-GR-entanglement}, and hence, is completely fixed by conformal symmetry\footnote{This is true for the perturbative entanglement entropy as well \cite{Rosenhaus:2014woa,Rosenhaus:2014ula,Allais:2014ata,Lewkowycz:2014jia,Rosenhaus:2014zza,Mezei:2014zla,Carmi:2015dla,Faulkner:2015csl,Leichenauer:2016rxw}.}. This \textit{universal} perturbative relative entropy for a general CFT was written in \cite{Faulk-GR-entanglement} as the symplectic flux of a scalar field in an auxiliary AdS-Rindler `bulk' spacetime %\tcb{(relation to JLMS)}%; see Eq.~\eqref{xx}. .
\begin{align}
S_{\text{rel}} (\rho||\sigma) \, = \,\, \int_{\Sigma_{0}} \,\,\, \omega_{\phi}\Big( \, \Phi\big(r , t , y \big) \, , \, \lie_{\xi} \Phi\big(r , t , y \big) \,  \Big) \,\, , \label{xx}
\end{align}
where $\xi$ is related to the timelike killing vector field and $\Sigma_{0}$ is a Cauchy slice in the AdS-Rindler spacetime; (see Sec.~(\ref{sec-bg-pert-re}) for details). On the other hand, when the excited state $\rho$ is not necessarily a perturbative state, the relative entropy between $\rho$ and $\sigma$ depends on the details of the CFT's. The relative entropy in this case was studied for holographic CFT's in \cite{lashkari2016gravitational}. It was argued that the bulk dual of the relative entropy is related to the difference in the conserved charges, $H_{\xi}\left(\MM_{\rho}\right)$ and $H_{\xi}\left(\MM_{\sigma}\right)$ , in the entanglement wedges of $\rho$ and $\sigma$ respectively. That is, \cite{lashkari2016gravitational}
\begin{align}
S_{\text{rel}}(\rho||\sigma) \, = \, H_{\xi}\left(\MM_{\rho}\right) - H_{\xi}\left(\MM_{\sigma}\right) \,\, . \label{yy}
\end{align}
The conserved charges $H_{\xi}$ have contributions from both the codimension-$2$ extremal surface and the asymptotic boundary.  We review the concept of conserved charges in Sec.~(\ref{sec-cons-charges}) and review the derivation of Eq.~\eqref{yy} in Sec.~(\ref{eq-non-pert-re-holo-bg}). 
%\newline

Another area of some recent interest has been that of studying the R\'enyi entropies, $S_n(\rho)$, defined by
\be
S_n(\rho)=\frac{1}{1-n}\log \tr \, \rho^{n} \, .
\ee
In particular, the holographic dual of a specific derivative of this quantity, called the refined R\'enyi entropy $\S_{n}(\rho)$, defined by
\be
\S_{n}(\rho)=n^2\partial_n\left(\frac{n-1}{n}S_n(\rho)\right) \,  \label{eq-Dong-Renyi}
\ee
was derived in \cite{Dong-Renyi}, where it was shown that this object is calculated by the minimal area of a cosmic brane with tension given by $T_n=\frac{n-1}{4nG_N}$. In the case when $n$ goes to one, this becomes the holographic formula for entanglement entropy \cite{RT, hubeny2007covariant,Lewkowycz:2013nqa,Dong:2016hjy} and corresponds to the tension-less limit of this brane. It is worth noting that while this refined R\'enyi entropy has a natural holographic dual, the R\'enyi entropy itself does not yet possess the same.
%\newline

Since the relative entropy is defined in the context of entanglement entropies, as, for example, is apparent from the fact that entanglement entropy appears explicitly in the form of the relative entropy, it is natural to ask for R\'enyi generalizations thereof. The problem is that the method for performing this generalization is not unique. Indeed, several different candidates have appeared in the literature, with some of the desirable properties that such a quantity should be expected to possess; see for example \cite{Wilde-14,sand13,FL-13,beigi,MO-15}. In particular, the quantity of the sandwiched R\'enyi relative entropy defined in these works seems to be the generalization that most preserves the desired properties, in particular the monotonicity of R\'enyi relative entropy under CPTP maps. This object is defined to be %[CITE]\tcr{I think we can delete this citation, as it's contained in the three above; delete this comment and the citation if you agree-Ning}
\be
S_{n} (\rho || \sigma) \, \equiv \, \frac{1}{n-1} \, \log\tr \left\{ \Big(\sigma^{\frac{1-n}{2n}} \, \rho \, \sigma^{\frac{1-n}{2n}} \Big)^{n} \right\} \, . \label{eq-srre-def-intro}
%S_n(\rho||\sigma)=\frac{1}{n-1}\log\Tr\left(\sigma^{\frac{1-n}{2n}}\rho\sigma^{\frac{1-n}{2n}}\right)^n.
\ee

It is now a natural question to ask whether such a generalization of the relative entropy can be nicely related to a geometric object with nice bulk properties via holography. The goal of this paper is to address this question. Specifically, we consider a refined version of the sandwiched R\'enyi relative entropy, which we define as
\begin{align}
\S_{n} (\rho||\sigma) \, \equiv \, n^{2} \, \partial_{n} \left( \frac{n-1}{n} \, S_{n} (\rho||\sigma) \right) \, . \label{eq-rrre-def-intro}
\end{align}
We call this quantity `refined' R\'enyi relative entropy. In particular, we study this quantity when the state $\sigma$ is the vacuum state of an arbitrary CFT reduced to a spherical region and $\rho$ is a slightly excited state. We find that, for integer $n \ge 1$, the perturbative refined R\'enyi relative entropy, just like the relative entropy, is completely fixed by conformal symmetry and can be written as the symplectic flux of a scalar field in an auxiliary AdS-Rindler spacetime. We therefore consider our result to be a one-parameter generalization of Eq.~\eqref{xx}. Moreover, we also specialize to the case of holographic CFT's and argue that the holographic dual of the non-perturbative refined R\'enyi relative entropy, for integer $n \ge 1$, is related to conserved charges in an asymptotically locally AdS spacetime. We consider this to be a one-parameter generalization of Eq.~\eqref{yy}. Finally, we use this holographic formula to prove certain `positive energy' theorems in the asymptotically locally AdS spacetimes. %Based on the fact that the refined R\'enyi relative entropy can be  
%\newline

It is important to point out here that the Hilbert space of a quantum field theory is not factorizable into subspaces of spatial subregions. Therefore, the `reduced' density matrices of a spatial subregion are not formally defined objects in quantum field theory. Moreover, local algebras in a quantum field theory are of Type III \cite{araki,Longo,Fredenhagen1985}, and therefore do not have a notion of trace. Hence, the definition of the relative entropy in Eq.~\eqref{eq-re-def-intro} and sandwiched R\'enyi relative entropy in Eq.~\eqref{eq-srre-def-intro} are only valid for finite dimensional Hilbert spaces. Nevertheless, the relative entropy between a general and a vacuum state on a subregion is still a well-defined object in quantum field theory. This is defined in terms of subalgebras of local operators in the subregion using Tometa-Takesaki modular theory \cite{zbMATH03511002,RCP25_1975__22__A1_0}; see \cite{Witten:2018lha} for review. Furthermore, a formal definition of the sandwiched R\'enyi relative entropy was recently discussed in \cite{Lashkari-sand}. Since, the sandwiched R\'enyi relative entropy is a well-defined quantity in a quantum field theory, we deduce from Eq.~\eqref{eq-rrre-def-intro} that the refined R\'enyi relative entropy is also a well-defined object in quantum field theory. 
%\newline

The rest of this paper in organized as follows. We start with a review of known results about relative entropy in CFT's, such as Eq.~\eqref{xx} and Eq.~\eqref{yy}, and a review of conserved charges using \textit{covariant phase space} methods in Sec.~(\ref{sec-bg}). We introduce refined R\'enyi relative entropy in Sec.~(\ref{sec-rrre-def}) and derive some properties of this quantity. In particular, we show that this quantity can be written as the relative entropy of an $n$-dependent state, which we call the `sandwiched' state. We use this observation to derive a general perturbative formula for the refined R\'enyi relative entropy in Sec.~(\ref{sec-gen-pert-rrre}). We then show in Sec.~(\ref{sec-setup}) that for a particular family of states, the sandwiched state can be prepared by a Euclidean path integral. Using these results, we study the refined R\'enyi {relative} entropy in Sec.~(\ref{sec-pert-analysis}) when the reference state is the vacuum of a CFT reduced to a spherical region and the other state is a small perturbation thereof. We show in Sec.~(\ref{sec-rrre-sf}) that the refined R\'enyi relative entropy in this case can be written as the symplectic flux of a scalar field through a Cauchy slice of the AdS-Rindler spacetime, thus providing a one-parameter generalization of Eq.~\eqref{xx}. We then specialize to CFT's with semi-classical holographic duals in Sec.~(\ref{sec-non-pert-analysis}). We argue that the holographic dual of the refined R\'enyi relative entropy between an excited state and a vacuum state reduced to a spherical region is related to conserved charges in the bulk dual of the sandwiched state. We therefore consider our result to be a one-parameter generalization of Eq.~\eqref{yy}. We use this result in Sec.~(\ref{sec-pos-energy-thm}) to prove certain `positive-energy' theorems in asymptotically locally AdS spacetimes. Finally, we end with a summary of our results and some possible future directions in Sec.~(\ref{sec-conc}).

%\tcr{REPHRASE THE FOLLOWING PARAGRAPH AFTER THE REST IS DONE TO MAKE IT A "IN THIS SECTION WE DO X" THING}

%In this work, we will define such a version of relative entropy built off of the refined relative entropy of \cite{Dong-Renyi} \tcb{(is this correct?)}, which we will call the refined R\'enyi relative entropy. We will then derive the properties of this quantity, and its holographic dual object. Finally, we will explore this object in the perturbative and non-perturbative limits, relating the former to a property of a symplectic form.

\section{Background} \label{sec-bg}

Our goal in this paper is to present a one-parameter generalization of the known results about the relative entropy in CFT's, particularly Eq.~\eqref{xx} and Eq.~\eqref{yy}. We dedicate this section to a brief review of these known results as well as a brief review of the concept of the conserved charges derived via covariant phase space methods.
%The concepts of relative entropy in conformal field theories and conserved charges in gravitational theories will be central to our analysis in this paper and we dedicate this section to a brief review of these concepts.

\subsection{Perturbative relative entropy in a general CFT} \label{sec-bg-pert-re}

As pointed out earlier, relative entropy has been studied in detail between states of a conformal field theory (CFT) in $R^{1,d-1}$ especially when the reference state is taken to be the vacuum state reduced to a spherical region. The simplification that occurs when dealing with the CFT vacuum reduced to a spherical region, $B$, is that the vacuum reduced state, $\sigma$, has a \textit{universal} form \cite{Casini:2011kv}. In particular, the modular Hamiltonian of the reduced vacuum state, defined as $K_{\sigma} \, \equiv \, - \log\sigma \,$ , is given by \cite{Casini:2011kv}
\begin{align}
K_{\sigma} \, = \, \int_{\mathcal{S}} \, d\Sigma^{b} \, \hat{\xi}^{a} \, T_{ab} \, , \label{eq-K-cov}
\end{align}
where $\mathcal{S}$ is any achronal slice within the domain of dependence of the spherical region $B$, $\mathcal{D}(B)$, such that $\partial \mathcal{S} \, = \, \partial B$. Moreover, the vector $\hat{\xi}$ is the generator of a (modular) flow in $\mathcal{D}(B)$. In particular, the vector field $\hat{\xi}$ (in Cartesian coordinates $x^{\mu} \, = \, \{x^{0},x^{i}\}$) for a spherical region of radius $R$ at $x^{0}=0$ and centered at the origin is 
\begin{align}
\hat{\xi} \, = \, \frac{\pi \, (R^{2} - (x^{0})^{2} - \x^{2}) }{R} \, \partial_{0} \, - \, \frac{2\pi \, x^{0} \, x^{i} }{R} \, \partial_{i} \, .
\end{align}
%\newline

For holographic CFT's on $R^{1,d-1}$, the vacuum state is dual to the Poincare patch of the $(d+1)$-dimensional vacuum AdS spacetime, which is described by
\begin{align}
ds^{2} \, = \, \frac{\ell^{2}}{z^{2}} \left( dz^{2} + \eta_{\mu\nu} dx^{\mu}dx^{\nu} \right) \, .
\end{align}
where $z \, = \, 0$ is the boundary of the AdS spacetime and $\ell$ is the AdS length scale. Moreover, the bulk region dual to the vacuum state reduced to a subregion is the corresponding entanglement wedge in Poincare AdS. The entanglement wedge corresponding to a spherical region $B$ of radius $R$, $\MM_{\sigma}$, is the domain of dependence of an achronal surface, $\Sigma_{0}$, such that $\partial\Sigma_{0} \, = \, B \cup \tilde{B}_{\sigma}$, where 
\begin{align}
\tilde{B}_{\sigma} \, : \, \quad x^{0} \, = \, 0 \, \quad \text{and} \, \quad  z^{2} + \x^{2} \, = \, R^{2} \, .
\end{align}
%where $z$ is the usual Fefferman-Graham radial coordinate [CITE]. 
The vector field, $\xi$, that generates the (modular) flow in the entanglement wedge $\MM_{\sigma}$ is 
\begin{align}
\xi^{z} \, = \, - \frac{2\pi \, x^{0} \, z}{R} \, \, \quad\quad \, \xi^{0} \, = \, \frac{\pi \, (R^{2} - z^{2} - (x^{0})^{2} - \x^{2}) }{R} \, \quad\quad \, \xi^{i} \, = \, \hat{\xi}^{i} \, .  \label{eq-eta-intro}
\end{align}
It is easy to check that $\xi$ is a killing vector field of the Poincare AdS spacetime and that it satisfies the following boundary conditions%. Furthermore, it vanishes at $\tilde{B}$ and it approaches $\hat{\eta}$ as $z \to 0$. 
\begin{align}
\xi\big|_{B} \, =& \,\, \hat{\xi} \, \quad\quad\quad\quad\quad\quad \xi\big|_{\tilde{B}_{\sigma}} \, = \, 0 \, . \label{eq-eta-bc-intro}
\end{align}
%\newline

The entanglement wedge $\MM_{\sigma}$ can be mapped, by a coordinate transformation, to the AdS-Rindler spacetime \cite{Casini:2011kv}, which is described by %\tcb{(add R)}
\begin{align}
ds^{2} \, = \, - \left(\frac{r^{2}}{R^{2}}-1\right) \, dt^{2} \, + \, \left(\frac{r^{2}}{R^{2}}-1\right)^{-1} dr^{2} \, + \, r^{2} \, ds^{2}_{\hp} \, , \label{eq-ads-rind-bg}
\end{align}
where $ds^{2}_{\hp}$ is the metric on the $(d-1)$-dimensional unit hyperbolic space. In this coordinate system, the extremal surface $\tilde{B}_{\sigma}$ is given by $r \, = \, R$ for any finite $t$, which  is the bifurcation surface of the AdS-Rindler spacetime. Moreover, $\xi$ is related to the time-translation vector field by %\tcb{check sign}
\begin{align}
\xi \, = \, -2\pi R \, \partial_{t} \, . \label{eq-eta-intro-2}
\end{align}
One can now check that the vector field $\xi$ satisfies
\begin{align}
\left( \nabla^{a}\xi^{b} \, - \nabla^{b}\xi^{a} \right)\big|_{\tilde{B}_{\sigma}} \, = \, 4\pi \, n^{ab} \, , \label{eq-binormal}
\end{align}
where $n^{ab} \, \equiv \, n_{1}^{a}n_{2}^{b} - n_{2}^{a}n_{1}^{b} \, ,$ where $n_{1} = \partial_{r}$ and $n_{2} = \partial_{t}$ are normal vectors to the extremal surface $\tilde{B}_{\sigma} \, $.
%\newline

Now let us consider an excited state of a CFT prepared by a path integral with the insertion of a smeared operator. We denote this state reduced to a spherical region $B$ by $\rho$. For the case when the excited state $\rho$ is a perturbation of $\sigma$, the relative entropy between these states for holographic CFT's was studied using the AdS/CFT correspondence in \cite{Lashkari:2015hha}. This work was generalized in \cite{Faulk-GR-entanglement} for all CFT's, irrespective of whether the CFT has a holographic dual or not. It was shown in \cite{Faulk-GR-entanglement} that the perturbative relative entropy can be written as the symplectic flux of a scalar field through a Cauchy slice of the AdS-Rindler spacetime. More precisely, 
\begin{align}
S_{\text{rel}} (\rho||\sigma) \, = \,\, \int_{\Sigma_{0}} \,\,\, \omega_{\phi}\Big( \, \Phi\big(r , t , y \big) \, , \, \lie_{\xi} \Phi\big(r , t , y \big) \,  \Big) \,\, . \label{xx-2}
\end{align}
The mass of the scalar field in Eq.~\eqref{xx-2} is fixed by the conformal dimension of the operator used to prepare the perturbed state $\rho$ whereas the boundary condition of the scalar field is fixed by the smearing function of that operator. If the CFT under consideration were holographic, this scalar field would have been the holographic dual of the operator used to prepare the state $\rho$. However, it should be noted that the AdS/CFT correspondence was not assumed in the analysis of \cite{Faulk-GR-entanglement} and hence, the result in Eq.~\eqref{xx-2} is valid for all CFT's. Therefore, the scalar field in Eq.~\eqref{xx-2} and the AdS-Rindler spacetime should be thought as auxiliary tools used to write the perturbed relative entropy geometrically.
%\newline

Recently, a similar perturbative calculation was performed in \cite{Ugajin:2018rwd} for the R\'enyi relative divergence which is defined as
\begin{align}
D_{n}(\rho||\sigma) \, \equiv \,  \tr \, \left( \sigma^{1-n} \, \rho^{n} \right) \, .
\end{align}
In particular, it was found that at the lowest order in the perturbation parameter, the quantity defined as
\begin{align}
\tilde{D}_{n}(\rho||\sigma) \, \equiv \, D_{-n}(\rho||\sigma) - D_{n}(\rho||\sigma) \, ,
\end{align}
and its derivative, $\partial_{n} \tilde{D}_{n}(\rho||\sigma) \,$, can also be written as the symplectic flux of the scalar field as in Eq.~\eqref{xx}. The perturbative R\'enyi relative divergence was also studied in \cite{add-ref-1}, where the R\'enyi relative divergence between two thermal states was written as a Euclidean path integral. This path integral was then computed using the AdS/CFT correspondence.
%\newline

 The sandwiched R\'enyi relative entropy and the R\'enyi relative divergence are both special cases of $n$-$z$ R\'enyi relative divergence which is defined as
\begin{align}
D_{n,z}(\rho||\sigma) \, \equiv \, \frac{1}{n-1} \, \log\tr \left\{ \Big(\sigma^{\frac{1-n}{2z}} \, \rho^{\frac{n}{z}} \, \sigma^{\frac{1-n}{2z}} \Big)^{z} \right\} \, .
\end{align}
This quantity was studied for two perturbatively nearby states in \cite{May:2018tir}. Indeed, it was found that at the lowest order in the perturbation parameter, this quantity can also be written as the symplectic flux of the scalar field as in Eq.~\eqref{xx}. However, the AdS/CFT correspondence was assumed in \cite{May:2018tir} and hence this analysis is only valid for holographic CFT's.

%For the case when $\rho$ is not a perturbation of $\sigma$, the relative entropy between these states for \textit{holographic} CFT's was studied in [CITE]. It was found that the holographic dual of the relative entropy between these states is related to certain conserved charges in the entanglement wedge corresponding to $\rho$, $\M_{\rho} \, $, and in the entanglement wedge corresponding to $\sigma$, $\M_{\sigma}\, $, according to [CITE] %. That is,
%\begin{align}
%   \S_{\text{rel}} (\rho||\sigma) \, = \,\, H_{\xi}\left(\M_{\rho}\right) - H_{\xi}\left(\M_{\sigma}\right) \,\, . \label{yy}
%\end{align}
%In this equation, $H_{\xi}\left(\M_{\sigma}\right)$ is the conserved charge \tcr{or hamiltonian} \tcr{(use hamiltonian (or both) instead because that is what is used in the next subsection?)} corresponding to the diffeomorphism generated by the killing vector field $\xi$ given in Eq.~\eqref{eq-eta-intro}. On the other hand, $H_{\xi}\left(\M_{\rho}\right)$ in the conserved charge corresponding to the diffeomorphism generated by any vector field in $\M_{\rho}$ that behaves like the vector field $\xi$ near the boundary of $\M_{\rho}$. That is, it is a killing vector field near the boundary and it satisfies boundary conditions similar to Eq.~\eqref{eq-eta-bc-intro}. %this vector field vanishes on the extremal surface corresponding to the state $\rho$ and this vector field approaches $\hat{\eta}$ as $z \to 0$.
%\newline

This concludes our review of the known perturbative results regarding relative entropy and various associated R\'enyi quantities. Before moving on to a review of the derivation of Eq.~\eqref{yy} presented in \cite{lashkari2016gravitational}, we review the notion of conserved charges and canonical energy in the following subsection.
%In the following subsection, \tcr{we review the notion of conserved charges in a theory of gravity, focusing in particular on the concept of canonical energy}. \tcb{(something like this.)}

\subsection{Quasi-local conserved charges and canonical energy} \label{sec-cons-charges} %{Conjugate Hamiltonian and canonical energy}

In this subsection, we switch from quantum information to gravity temporarily and talk about the concepts of conserved charges and canonical energy in the general framework of covariant phase space methods. %The holographic dual of the relative entropy is related to the conserved charges as given in Eq.~\eqref{yy}  [CITE]. %We will see in the following section that this turns out to be related the holographic dual of refined R\'enyi relative entropy.
We restrict ourselves to a review of these quantities only to the extent that will be relevant for our analysis in this paper. The reader is referred to canonical references on this topic \cite{W-noether-entropy,IW-noether-entropy,LW, WZ} for full details.
%\newline

%A particularly useful way to study asymptotic symmetries and corresponding charges in the context of gravity is using covariant phase space methods, first developed in \cite{W-noether-entropy,IW-noether-entropy,LW}. Here, one defines a configuration space of all kinematically allowed field configurations of a theory, for example, the space of possible metrics for the case of pure gravity, here denoted by $\mathcal{G}$ where the space of on-shell field configurations is a subspace of $\mathcal{G}$, denoted by $\bar{\mathcal{G}}$. Different points in the configuration space are connected by perturbations in the fields. In particular, perturbations that interpolate between points in $\bar{\mathcal{G}}$, must satisfy linearized the linearized equations of motion of the theory in consideration. 
Given a field theory in $(d+1)$ dimensions, any variation of the Lagrangian is given by
\be\label{lage}
\delta L \equiv E(g) \delta g + d\theta (g,\delta g)\,,
\ee
\noindent
where we use $g$, $\delta g$ as collective labels  to represent the metric and matter fields and their variations respectively. In Eq.~\eqref{lage}, $E(g) = 0$ denotes the equations of motion and the $d$-form $\theta(g,\delta g)$ is called the {symplectic potential}. %\tcr{Depending on who's reading, the distinction between presymplectic and symplectic that I wrote here is important. I am not sure if we should keep it or discard it though.} \tcb{($\Theta$ and $\omega$ are symplectic object. You may have to worry about this distinction for $\Omega \equiv \int \omega$, but we are not talking about that here. So then why get into the discussion of presymplectic at all? ) \tcr{(They are not symplectic objects. In gauge theories, they are degenerate and are called in general presymplectic. A symplectic manifold is defined by a non-degenerate form. You mod out by degeneracies, make $\Omega$ non-degenerate, that gives you Poisson brackets and then these things are called symplectic. Having said that, pointing out this distinction may just be a matter of taste so I wouldn't insist on it.)} }.
%\footnote{Technically, one only obtains \textit{symplectic} quantities from \textit{presymplectic} quantities after taking a symplectic quotient and modding out by degeneracies \cite{LW}. After this, the field configuration space acquires the structure of a symplectic manifold. The submanifold that corresponds to on-shell field configurations is then called the \textit{covariant phase space}. This distinction, however, will not concern us here and so we will stick to using the term \textit{symplectic} everywhere instead of \textit{presymplectic}.}.
From the symplectic potential, one can define the {symplectic current}, $\omega(g,\delta_{1}g,\delta_{2}g)$, as 

\be\label{sympcurrent}
\omega(g,\delta_{1}g,\delta_{2}g) \equiv \delta_{1} \theta (g,\delta_{2} g) - \delta_{2} \theta(g,\delta_{1} g) \, ,
\ee
which is a $d$-form and is antisymmetric and bilinear in two field perturbations. %This allows us to define what is called the \textit{perturbed} hamiltonian conjugate to $\eta^{a}$, which is given by:
%\newline

Given $\Sigma$, a subspace of a codimension-1 %\tcr{(doesn't have to be spacelike, most of this stuff is applied to null surfaces anyway)} 
hypersurface and $\eta$, a vector field on $\Sigma$, the \textit{perturbed} Hamiltonian conjugate to $\eta$ is given by \cite{W-noether-entropy,IW-noether-entropy,LW, WZ}%[cite]
\begin{equation} \label{dive}
\delta H_{\eta} \, \equiv \, \int_{\Sigma} \, \omega( g , \delta g, \lie_{\eta}g) \, .
\end{equation}
Using Eq.~\eqref{lage} and assuming equations of motions, $E(g) \, = \, 0$, this can be written as
\begin{align}
\delta H_{\eta} \, = \, \int_{\Sigma} \, \delta \Big( \theta(\lie_{\eta}g) - \eta \cdot L \Big) \, - \, \int_{\partial\Sigma} \, \eta\cdot\theta(\delta g) \, , \label{eq-var-h}
\end{align}
where $\partial\Sigma$ denotes the boundary of $\Sigma$.
%Finally, the Noether two form, in the case of Einstein-Hilbert gravity minimally coupled to matter, is given by
% \be
%Q_{ab}[\eta] = \frac{-1}{16 \pi} \epsilon_{abcd} \nabla^{c} \eta^{d} \,,
%\ee
% 
%where $\delta Q_{\eta}$ is called the \textit{perturbed} Noether form. 
It is worth pointing out that the existence of the Hamiltonian, $H_{\eta}$, is a non-trivial issue. In particular, it exists if and only if the boundary term is a total variation which is to say that the quantity appearing on the right hand side of Eq.~\eqref{eq-var-h} is integrable. This is guaranteed if the following condition holds \cite{WZ}
\begin{align}
\int_{\partial\Sigma} \, \eta \cdot \left[ \delta_{1}\theta(\delta_{2}g) - \delta_{2}\theta(\delta_{1}g) \right] \, = \, 0 \, , \label{eq-bc}
\end{align} 
for arbitrary perturbations $\delta_{1}$ and $\delta_{2}$. When this condition holds, there exists a $d$-form $K(g)$ such that  \cite{lashkari2016gravitational}
\begin{align}
\delta \big(\eta\cdot K(g)\big) \, \equiv \, \eta \cdot \theta(\delta g) \,\, \quad\quad \text{on } \,\,\,\, \partial\Sigma \, .  \label{eq-def-K}
\end{align}

Assuming the condition in Eq.~\eqref{eq-bc} is satisfied, the Hamiltonian conjugate to $\eta$ can be written as
\begin{align}
H_{\eta} \, = \, \int_{\Sigma} \, J_{\eta} \, - \, \int_{\partial\Sigma} \, \eta \cdot K \, ,  \label{eq-def-H}
\end{align}
where
\begin{align}
J_{\eta} \, \equiv \, \theta(\lie_{\eta}g) - \eta \cdot L \, . \label{eq-def-J}
\end{align}
%\newline

Note that the Hamiltonian, $H_{\eta}$, in Eq.~\eqref{eq-def-H} can be written purely as a boundary integral. This is because $J_{\eta}$ can be written, using equations of motion, as 
\begin{align}
J_{\eta} \, = \, d Q_{\eta} \, , \label{eq-charge}
\end{align}
where $Q_{\eta}$ is a $(d-1)$-form and is called the Noether charge. Therefore, the Hamiltonian can be written as an integral over $\partial\Sigma$,
\begin{align}
H_{\eta} \, = \, \int_{\partial\Sigma} \, \left( Q_{\eta} - \eta \cdot K \right) \, . \label{eq-can-ener-charge}
\end{align}
%\newline
This `quasi-local' expression implies that the Hamiltonian only depends on the details of the vector field $\eta$ near the boundary $\partial\Sigma$. Moreover, this implies that the Hamiltonian is the same for any two codimension-$1$ surfaces $\Sigma_{1}$ and $\Sigma_{2}$ if they have a common boundary, \textit{i.e.}   $\partial\Sigma_{1} \, = \, \partial\Sigma_{2} \, $. In this sense, the Hamiltonian is conserved and hence we will use it interchangeably with \textit{conserved charge} in this paper. %This fact will play an important role in Sec.~(\ref{sec-non-pert-analysis}).
%\newline

%In this paper, we will only focus on Einstein gravity minimally coupled to matter with \tcr{$G_{N}$ set to 1}. In this case, the Noether charge is given by \cite{IW-noether-entropy} 
%\be
%Q_{\eta_{a_{3}\cdots a_{d-1}}} \, = \, - \, \frac{1}{16 \pi} \tilde{\epsilon}_{a_{1}a_{2} a_{3} \cdots a_{d-1}} \nabla^{a_{1}} \eta^{a_{2}} \, ,
%\ee
%where $\tilde{\epsilon}$ denotes the $(d+1)$ dimensional volume form.
%\newline

Now let us suppose that the metric and matter fields have a perturbative expansion given by
\begin{align}
g \, = \, g^{(0)} \, + \, \epsilon \, g^{(1)} \, + \, O(\epsilon^{2}) \, , \label{eq-met-pert}
\end{align}
where $\epsilon \ll 1$ is a perturbation parameter. Let us further assume that the unperturbed metric and matter fields are invariant under the diffeomorphism generated by the vector field $\eta$. That is, $\lie_{\eta} \, g^{(0)} \, = \, 0 \, $. Using Eq.~\eqref{dive}, we deduce that the derivative of the Hamiltonian, $H_{\eta} \, $, is given by
\begin{align}
\frac{d}{d\epsilon} H_{\eta} \, = \, \int_{\Sigma} \, \omega\left( g , \frac{d}{d\epsilon} g, \lie_{\eta}g \right) \, .
\end{align}
At linear order in $\epsilon$, the Hamiltonian, $H_{\eta} \, $, is %\tcr{(Can we remove this equation? It is confusing, it makes it seem like the epsilon variation only happens in the first argument.)}\textcolor{blue}{(But this is precisely what is happening.)}
\begin{align}
H_{\eta}^{(1)} \, \equiv \, \frac{d H_{\eta}}{d\epsilon} \Big|_{\epsilon = 0} \, = \, \int_{\Sigma} \, \omega\left( g^{(0)}, g^{(1)}, \lie_{\eta} g^{(0)} \right) \, . \label{eq-pert-Heta-one} %\, \int_{\Sigma} \, \omega\left( g , \frac{d}{d\epsilon} g, \lie_{\eta}g \right)\Bigg|_{\epsilon = 0} \, .
\end{align}
Since we have assumed %$\eta$ is assumed to a killing vector field of the unperturbed metric, we have
$\lie_{\eta} g^{(0)} \, = \, 0 \, $, we get that $H_{\eta}^{(1)} \, = \, 0 \, $. Similarly, the Hamiltonian, $H_{\eta} \, $, at quadratic order in $\epsilon$ is given by %\tcr{If $g$ here represents both the metric and the scalar field, don't we have to clarify what is going on with the perturbations to the scalar field? This would clarify the connection with what's being done in Sec. (5.2)?}
%\tcr{(I just mean that the line below Eq. 37 only says what happens to the lie derivative of the zeroth order. If g is a generic label for the metric and scalar field, don't we need to explicitly say what happens to the scalar field?)}\tcb{same things happens to background scalar field i.e. its invariant under eta. i agree saying eta is killing field is not enough. was planning to fix it at the end.}
\begin{align}
H_{\eta}^{(2)} \, \equiv \, \frac{d^{2} H_{\eta}}{d\epsilon^{2}} \Big|_{\epsilon = 0} \, = \, \int_{\Sigma} \, \omega\left( g^{(0)}, g^{(1)}, \lie_{\eta} g^{(1)} \right) \, . \label{eq-pert-Heta-two} %\, \int_{\Sigma} \, \omega\left( g , \frac{d}{d\epsilon} g, \lie_{\eta}g \right)\Bigg|_{\epsilon = 0} \, .
\end{align}
The quantity on the right hand side, for the case when $\eta$ is a timelike killing vector field of the unperturbed metric, %the vector field corresponds to time translations} 
is called the \textit{canonical energy} \cite{HW}. %\tcr{(I thought it was called canonical energy only if this vector was a (asymtotically a time translation but maybe this is just semantics)}\textcolor{blue}{(I take my previous comment back. I think I agree with you. But it is enough to call $\eta$ to be  a timelike vector as it is already taken to be killing vector.) }. [cite: Hollands and Wald]. {\tcr{(Why second order? Why isn't the full thing canonical energy?)} \textcolor{blue}{I do not know. This is how Hollands-Wald have defined it. They were studying stability (of BH, I guess?) after perturbations. Moreover, this is precisely what Lashkari-Van Raamsdonk and Faulkner-etal have in their paper.}}  %\tcr{associated to the region $\Sigma$.}
%\newline

%In the particular case where the vector field $\eta$ corresponds to an asymptotic time translation, this hamiltonian is referred to as the canonical energy of the region bounded by $\delta \Sigma$. 
This concludes our review of the concepts of conserved charges and canonical energy. In the next subsection, we review the derivation of Eq.~\eqref{yy} originally presented in \cite{lashkari2016gravitational}. 

\subsection{Relative entropy in holographic CFT's} \label{eq-non-pert-re-holo-bg}

Now we consider a CFT with a holographic dual. Like in Sec.~(\ref{sec-bg-pert-re}), we denote an excited state and a vacuum state reduced to a spherical region by $\rho$ and $\sigma$ respectively. However, we no longer assume that the state $\rho$ is a perturbative state around the vacuum state $\sigma$. The relative entropy between these states was studied in \cite{lashkari2016gravitational} using the AdS/CFT correspondence. It was found that the holographic dual of the relative entropy between these states is related to certain conserved charges in the entanglement wedge corresponding to $\rho$, $\MM_{\rho} \, $, and in the entanglement wedge corresponding to $\sigma$, $\MM_{\sigma}\, $. %, according to [CITE] 
%\begin{align}
%    \S_{\text{rel}} (\rho||\sigma) \, = \,\, H_{\xi}\left(\M_{\rho}\right) - H_{\xi}\left(\M_{\sigma}\right) \,\, . \label{yy}
%\end{align}
Here, we review the argument of \cite{lashkari2016gravitational}. 
%\newline

Recall that the relative entropy in Eq.~\eqref{eq-re-def-intro} can be written as
\begin{align}
S_{\text{rel}}(\rho||\sigma) \, = \, S(\sigma) \, - \, S(\rho) \, + \, \langle K_{\sigma} \rangle_{\rho} \, - \, \langle K_{\sigma} \rangle_{\sigma} \, , \label{eq-s-rel-nonpert-bg}
\end{align}
where $S(\rho)$ and $S(\sigma)$ are the  von Neumann entropies of the states $\rho$ and $\sigma$ whereas $\langle K_{\sigma}\rangle_{\rho}$ and $\langle K_{\sigma}\rangle_{\sigma}$ are the expectation values of $K_{\sigma}$ in the states $\rho$ and $\sigma$ respectively. 
%\newline

For holographic CFT's, we already know the bulk duals of $S(\sigma)$ and of $S(\rho)$. In particular, these quantities are given by \cite{RT,hubeny2007covariant}
\begin{align}
S(\sigma) \, =& \,\, \frac{\text{Area}\big(\tilde{B}_{\sigma}\big)}{4 \, G_{N}} \, , \label{eq-s-sigma-rt} \\
S(\rho) \, =& \,\, \frac{\text{Area}\big(\tilde{B}_{\rho}\big)}{4 \, G_{N}} \, , \label{eq-s-rho-rt}
\end{align}
where $\tilde{B}_{\sigma}$ and $\tilde{B}_{\rho}$ are the boundary anchored codimension-$2$ extremal surface corresponding to the states $\sigma$ and $\rho$ respectively. Now using the form of $K_{\sigma}$ from Eq.~\eqref{eq-K-cov} and using Eq.~\eqref{eq-s-sigma-rt} and Eq.~\eqref{eq-s-rho-rt}, we write the relative entropy in Eq.~\eqref{eq-s-rel-nonpert-bg} as
\begin{align}
S_{\text{rel}}(\rho||\sigma) \, = \, \frac{\text{Area}\big(\tilde{B}_{\sigma}\big)}{4 \, G_{N}} \, - \, \frac{\text{Area}\big(\tilde{B}_{\rho}\big)}{4 \, G_{N}} \, + \, \int_{\mathcal{S}} \, d\Sigma^{b} \, \hat{\xi}^{a} \, \langle T_{ab} \rangle_{\rho} \, - \, \int_{\mathcal{S}} \, d\Sigma^{b} \, \hat{\xi}^{a} \, \langle T_{ab} \rangle_{\sigma} \, . \label{eq-s-rel-nonpert-bg-two}
\end{align}
%\newline

As discussed in Sec.~(\ref{sec-bg-pert-re}), $\tilde{B}_{\sigma}$ is the bifurcation surface of the AdS-Rindler spacetime. Hence, the area of this surface can be written as a surface integral of the Noether charge conjugate to $\xi$ \cite{W-noether-entropy,IW-noether-entropy}. That is,
\begin{align}
\frac{\text{Area}\big(\tilde{B}_{\sigma}\big)}{4 \, G_{N}} \, = \, \int_{\tilde{B}_{\sigma}} \, Q_{\xi}\left(\MM_{\sigma}\right) \, , \label{eq-area-sig-nc}
\end{align}
where $\xi$ is the killing vector field in Eq.~\eqref{eq-eta-intro}. Moreover, since $\xi$ vanishes at $\tilde{B}_{\sigma}$ as stated in Eq.~\eqref{eq-eta-bc-intro}, we can write Eq.~\eqref{eq-area-sig-nc} as
\begin{align}
\frac{\text{Area}\big(\tilde{B}_{\sigma}\big)}{4 \, G_{N}} \, = \, \int_{\tilde{B}_{\sigma}} \, \left( Q_{\xi}\big(\MM_{\sigma}\big) \, - \, \xi \cdot K\big(\MM_{\sigma}\big) \right) \, . \label{eq-area-sig-nc-two}
\end{align}
%\newline

In general, there is no killing vector field in the bulk dual of an excited CFT state. Therefore, it does not trivially follow that the area of $\tilde{B}_{\rho}$ can also be written as in Eq.~\eqref{eq-area-sig-nc-two}. However, it was argued in \cite{lashkari2016gravitational} that one can always find a vector field (which we also denote by $\xi$) that vanishes on  $\tilde{B}_{\rho}$ and that satisfies Eq.~\eqref{eq-binormal} near $\tilde{B}_{\rho}$. With these conditions on $\xi$, it was shown in \cite{lashkari2016gravitational} that we can write
\begin{align}
\frac{\text{Area}\big(\tilde{B}_{\rho}\big)}{4 \, G_{N}} \, = \, \int_{\tilde{B}_{\rho}} \, \left( Q_{\xi}\big(\MM_{\rho}\big) \, - \, \xi \cdot K\big(\MM_{\rho}\big) \right) \, . \label{eq-area-rho-nc-two}
\end{align}
%\newline

If the state $\rho$ is a perturbative state around the vacuum, then it was shown in \cite{Faulkner:2013ica} that
\begin{align}
\int_{B} \, \left( Q_{\xi}\big(\MM_{\rho}\big) \, - \, \xi \cdot K\big(\MM_{\rho}\big) \right) \, - \, \int_{B}& \, \left( Q_{\xi}\big(\MM_{\sigma}\big) \, - \, \xi \cdot K\big(\MM_{\sigma}\big) \right) \,\nonumber\\ =& \, \int_{\mathcal{S}} \, d\Sigma^{b} \, \hat{\xi}^{a} \, \langle T_{ab} \rangle_{\rho} \, - \, \int_{\mathcal{S}} \, d\Sigma^{b} \, \hat{\xi}^{a} \, \langle T_{ab} \rangle_{\sigma} \, , \label{eq-ener-non-pert-bg}
\end{align}
where the integral is performed over a spherical boundary region $B$. It was then argued in \cite{lashkari2016gravitational} that this expression holds even if $\rho$ is not a perturbative state. 
%\newline

Now using Eq.~\eqref{eq-area-sig-nc-two}, Eq.~\eqref{eq-area-rho-nc-two}, and Eq.~\eqref{eq-ener-non-pert-bg}, we write Eq.~\eqref{eq-s-rel-nonpert-bg-two} as
\begin{align}
S_{\text{rel}}(\rho||\sigma) \, = \,\, &\int_{B} \, \left( Q_{\xi}\big(\MM_{\rho}\big) \, - \, \xi \cdot K\big(\MM_{\rho}\big) \right) \, - \, \int_{\tilde{B}_{\rho}} \, \left( Q_{\xi}\big(\MM_{\rho}\big) \, - \, \xi \cdot K\big(\MM_{\rho}\big) \right) \, \nonumber\\
- \, &\int_{B} \, \left( Q_{\xi}\big(\MM_{\sigma}\big) \, - \, \xi \cdot K\big(\MM_{\sigma}\big) \right) \, + \, \int_{\tilde{B}_{\sigma}} \, \left( Q_{\xi}\big(\MM_{\sigma}\big) \, - \, \xi \cdot K\big(\MM_{\sigma}\big) \right) \, . \label{eq-s-rel-holo-ints}
\end{align}
Note that $B \cup \tilde{B}_{\sigma}$ and $B \cup \tilde{B}_{\rho}$ are the boundaries of achronal slices on $\MM_{\sigma}$ and $\MM_{\rho}$ respectively. Now using Eq.~\eqref{eq-can-ener-charge}, we write the relative entropy in Eq.~\eqref{eq-s-rel-holo-ints} as
\begin{align}
S_{\text{rel}}(\rho||\sigma) \, = \, H_{\xi}\left(\MM_{\rho}\right) - H_{\xi}\left(\MM_{\sigma}\right) \,\, . \label{yy-2}
\end{align}
In this equation, $H_{\xi}\left(\MM_{\sigma}\right)$ is the conserved charge corresponding to the diffeomorphism generated in $\MM_{\sigma}$ by the killing vector field $\xi$ given in Eq.~\eqref{eq-eta-intro}. On the other hand, $H_{\xi}\left(\MM_{\rho}\right)$ is the conserved charge corresponding to the diffeomorphism generated by any vector field in $\MM_{\rho}$ that behaves like the vector field $\xi$ near the boundary of $\MM_{\rho}$. That is, it is a killing vector field near the asymptotic boundary and it satisfies boundary conditions similar to Eq.~\eqref{eq-eta-bc-intro} and Eq.~\eqref{eq-binormal}.
%\newline

This finishes our review of the known results about relative entropy in the context of CFT's and holography, especially Eq.~\eqref{xx} and Eq.~\eqref{yy}. Our goal in this paper is to show that similar results are true for the refined R\'enyi relative entropy defined in Eq.~\eqref{eq-rrre-def-intro}. Before we do this, we discuss the refined R\'enyi relative entropy and derive some of its properties in the next section. 

\section{Preliminaries} 

\subsection{Refined R\'enyi relative entropy} \label{sec-rrre-def}

We start by reviewing two quantities in quantum information theory that are particularly relevant to this paper, namely relative entropy and R\'enyi relative entropy. In doing so, we will show that for particular choices of states, relative entropy and a quantity related to R\'enyi relative entropy, the \textit{refined} R\'enyi relative entropy, are actually equal to each other. This equality will be central to our analysis in the subsequent sections of this paper.
%\newline

Recall that the relative entropy between two states $\rho$ and $\sigma$ is defined as
\begin{align}
S_{\text{rel}}(\rho||\sigma) \, \equiv \, \tr( \rho\log\rho ) - \tr (\rho\log\sigma) \, , \label{eq-re-def}
\end{align}
when supp($\rho$) $\subseteq $ supp($\sigma$) (otherwise, the relative entropy is taken to be $\infty$). The relative entropy satisfies a data-processing inequality which implies that the relative entropy decreases under a completely positive and trace preserving map (CPTP) \cite{Lindblad1974}. That is, under a CPTP map $\mathcal{E}$, 
\begin{align}
S_{\text{rel}}\left(\mathcal{E}(\rho)||\mathcal{E}(\sigma)\right) \, \le \, S_{\text{rel}}(\rho||\sigma) \, . \label{eq-dp-rel}
\end{align}
Now consider two reduced density states of a subsystem $A$, $\rho_{A}$ and $\sigma_{A}$ . For any smaller subsystem $B \subset A$, Eq.~\eqref{eq-dp-rel} implies
\begin{align}
S_{\text{rel}}\left(\rho_{B}||\sigma_{B}\right) \, \le \, S_{\text{rel}}(\rho_{A}||\sigma_{A}) \, . \label{eq-mono-rel}
\end{align}
In other words, it is more difficult to distinguish two states when some potentially distinguishing information is traced out.%we are given the states on a smaller subsystem \tcr{(better words)}.
%\newline

The one-parameter generalization of the relative entropy that also satisfies a data-processing inequality is the `sandwiched' R\'enyi relative entropy, which is defined as \cite{Wilde-14,sand13,FL-13,beigi,MO-15}%\cite{FL-13, Wilde-14,MO-15,beigi,sand13 }%[cite]
\begin{align}
S_{n} (\rho || \sigma) \, \equiv \, \frac{1}{n-1} \, \log\tr \left\{ \Big(\sigma^{\frac{1-n}{2n}} \, \rho \, \sigma^{\frac{1-n}{2n}} \Big)^{n} \right\} \, ,  \label{eq-ren-rel}
\end{align}
when supp($\rho$) $\subseteq $ supp($\sigma$). This quantity also monotonically increases with R\'enyi parameter $n$ \cite{sand13,beigi}. That is,
%In addition to being monotonic under CPTP maps (data-processing inequality) [cite], the sandwiched R\'enyi relative entropy are also monotonic in $n$ [cite]. In particular,
\begin{align}
\partial_{n} \, S_{n} (\rho || \sigma) \, \ge \, 0 \, . \label{eq-mono-sand}
\end{align}
%\newline

In this work, we introduce a closely related quantity, \textit{refined} R\'enyi relative entropy, which is defined as
\begin{align}
\S_{n} (\rho||\sigma) \, \equiv \, n^{2} \, \partial_{n} \left( \frac{n-1}{n} \, S_{n} (\rho||\sigma) \right) \, . \label{eq-mod-ren}
\end{align}
Just like the sandwiched relative entropy, this quantity approaches the relative entropy between states $\rho$ and $\sigma$ in the limit $n \to 1$. That is,
\begin{align}
\lim_{n\to 1} \S_{n}(\rho||\sigma)  \, = \, S_{\text{rel}}(\rho||\sigma) \, .
\end{align}
Note that we can write Eq.~\eqref{eq-mod-ren} as
\begin{align}
\S_{n} (\rho||\sigma) \, = \, S_{n} (\rho||\sigma)  \, + \, n(n-1) \, \partial_{n}S_{n} (\rho||\sigma) \, .
\end{align}
Now using the monotonicity of sandwiched R\'enyi relative entropy, Eq.~\eqref{eq-mono-sand}, we find the following hierarchical structure between the different relative entropies
\begin{align}
\S_{n} (\rho||\sigma) \, \ge \, S_{n} (\rho||\sigma) \, \ge \,  S_{\text{rel}}\left(\rho || \sigma\right) \, \quad\quad \text{ for } n \, \ge \, 1 \, , \label{eq-RRRE-ineq}\\
\S_{n} (\rho||\sigma) \, \le \, S_{n} (\rho||\sigma)  \, \le \,  S_{\text{rel}}\left(\rho || \sigma\right) \quad\quad \text{ for } n \, \le \, 1 \, . \label{eq-RRRE-ineq-2}
\end{align}
As we will see in Sec.~(\ref{sec-pos-energy-thm}), applying the inequality in Eq.~\eqref{eq-RRRE-ineq} in the context of the AdS/CFT correspondence leads to interesting energy conditions in asymptotically AdS spacetimes. 
%\newline

A useful property of the refined R\'enyi relative entropy is that it can be written as the relative entropy of a related state. More precisely, we have
\begin{align}
\S_{n} (\rho||\sigma) \, =& \,\, S_{\text{rel}} \left(\r||\sigma\right) \, ,           \label{eq-ren-rel-fin}
\end{align}
where we have introduced a `sandwiched' state, $\r$, which is defined as 
\begin{align}
\r \, \equiv \, \frac{ \, \Big({\sigma^{\frac{1-n}{2n}} \, \rho \, \sigma^{\frac{1-n}{2n}}}\Big)^{n} \, }{\tr \, \Big({\sigma^{\frac{1-n}{2n}} \, \rho \, \sigma^{\frac{1-n}{2n}}}\Big)^{n} \,} \, .  \label{eq-sand-st}
\end{align}
%where  
%\begin{align}
%\hat{\rho}_{(n)} \, \equiv \, \Big({\sigma^{\frac{1-n}{2n}} \, \rho \, \sigma^{\frac{1-n}{2n}}}\Big)^{n} \, . \label{eq-sand-int}
%\end{align}
We relegate the derivation of Eq.~\eqref{eq-ren-rel-fin} to Appendix~(\ref{app-sn}) and discuss the consequences of this identity in the following. First, the fact that the relative entropy is non-negative guarantees that the refined R\'enyi relative entropy is also non-negative. 
\begin{align}
\S_{n} (\rho||\sigma) \, \ge \, 0 \, .
\end{align}
Secondly, the identity in Eq.~\eqref{eq-ren-rel-fin} allows us to study the perturbative expansion of the refined R\'enyi relative entropy which we discuss in detail in the next subsection. 
%\newline

\subsubsection{Perturbation of the refined R\'enyi relative entropy} \label{sec-gen-pert-rrre}

In this section, our goal is to study the refined R\'enyi relative entropy when  the state $\rho$ is perturbatively expanded around an arbitrary state $\sigma$. More precisely, we take state $\rho$ to be given by
\begin{align}
\rho \, = \, \sigma \, + \, \epsilon \, \rho^{(1)} \, + \, O(\epsilon^{2}) \, , \label{eq-rho-ser}
\end{align}
where $\epsilon \ll 1$ is a perturbation parameter. The perturbation of the relative entropy has been studied in the literature \cite{Faulk-GR-entanglement,Faulkner:2014jva,Lashkari:2015hha}. Since the relative entropy is non-negative, it means that the relative entropy between states $\rho$ and $\sigma$ vanishes at linear order in $\epsilon$. This implies that the change in entanglement entropy is equal to the change in `modular' energy, which is usually called the first law of entanglement \cite{Blanco:2013joa}. Therefore, to lowest order in $\epsilon$, the relative entropy between states $\rho$ and $\sigma$ is of the form 
\begin{align}
S_{\text{rel}}(\rho||\sigma) \, = \,  \frac{\epsilon^{2}}{2} \, S^{(2)}_{\text{rel}}(\rho||\sigma) \, + \, O(\epsilon^{3}) \, ,
\end{align}
where \cite{Faulk-GR-entanglement}
\begin{align}
S^{(2)}_{\text{rel}}(\rho||\sigma) \, = \, - \, \int_{-\infty}^{\infty} \, \frac{ds}{4 \, \sinh^{2}\left(\frac{s+i\delta}{2}\right)} \,\,\, \tr \Big( \sigma^{-1 - \frac{is}{2\pi}} \, \rho^{(1)} \, \sigma^{\frac{is}{2\pi}} \, \rho^{(1)} \Big) \, . \label{eq-s-rel-pert} 
\end{align}
%\newline

Since the refined R\'enyi relative entropy is related to the relative entropy according to Eq.~\eqref{eq-ren-rel-fin}, we can deduce the second-order contribution to the refined R\'enyi relative entropy using Eq.~\eqref{eq-s-rel-pert}. To do this, we first need to find the perturbative expansion of the sandwiched state, $\r$. By inserting the expansion of $\rho$ from Eq.~\eqref{eq-rho-ser} in Eq.~\eqref{eq-sand-st}, we find that the sandwiched state for integer $n \ge 1$ at linear order in $\epsilon$ is given by
\begin{align}
\r \, = \, \sigma \, + \, \epsilon \, \sum_{k = 0}^{n-1} \, \, \sigma^{\frac{1-n+2k}{2n}} \, \rho^{(1)} \, \sigma^{\frac{n-1-2k}{2n}} \, + O(\epsilon^{2}) \, .
\end{align}
Now using Eq.~\eqref{eq-ren-rel-fin} and Eq.~\eqref{eq-s-rel-pert}, we deduce that the refined R\'enyi relative entropy at the lowest order in $\epsilon$ is 
\begin{align}
\S_{n}(\rho||\sigma) \, = \,  \frac{\epsilon^{2}}{2} \, \S^{(2)}_{n}(\rho||\sigma) \, + \, O(\epsilon^{3}) \, , \label{eq-sn-ser-exp}
\end{align}
where
\begin{align}
\S^{(2)}_{n} (\rho||\sigma) \, = \, - \, \,\sum_{k=0}^{n-1}\sum_{j=0}^{n-1} \,\,  \int_{-\infty}^{\infty} \, \frac{ds}{4 \,\sinh^{2}\left(\frac{s+i\delta}{2}\right)} \,\,\, \tr \Big( \sigma^{-1} \,\, \sigma^{-\frac{is}{2\pi}+\frac{(k-j)}{n}} \, \rho^{(1)} \, \sigma^{\frac{is}{2\pi}-\frac{(k-j)}{n}} \, \rho^{(1)} \Big) \, . \label{eq-ren-rel-pert-fin}
\end{align}
This general formula for the perturbation of the refined R\'enyi relative entropy for integer $n\ge 1$ is the main result of this section\footnote{It may be possible to analytically continue Eq.~\eqref{eq-ren-rel-pert-fin} to non-integer $n$ by converting the sum into a contour integral as done in \cite{Faulkner:2014jva}. We thank Tom Hartman for pointing this out.}. We will use this result in Sec.~(\ref{sec-pert-analysis}) to relate the perturbative refined R\'enyi relative entropy to the symplectic flux of a scalar field through a Cauchy slice of the AdS-Rindler wedge. %.... \tcr{(include a one-line summary of the perturbative result)}.

%Now let's consider a `state-dependent' CPTP map on a space of density matrices defined as
%\begin{align}
%    \mathcal{E}_{\sigma} \, : \, \xi \, \to \,  \frac{\Big(\sigma^{\frac{1-n}{2n}} \, \xi \, \sigma^{\frac{1-n}{2n}} \Big)^{n}}{\tr \Big(\sigma^{\frac{1-n}{2n}} \, \xi \, \sigma^{\frac{1-n}{2n}} \Big)^{n}} \, .
%\end{align}
%Using this, we can write Eq.~\eqref{eq-ren-rel-fin} as
%\begin{align}
%    \S_{n} (\rho||\sigma) \, =& \,\, S_{\text{rel}} \left(\mathcal{E}_{\sigma}(\rho)||\mathcal{E}_{\sigma}(\sigma)\right) \, .
%\end{align}
%Now using data-processing inequality, Eq.~\eqref{eq-dp-rel}, we deduce 
%\begin{align}
%    \S_{n} (\rho||\sigma) \, \le \,\, S_{\text{rel}} \left(\rho||\sigma\right) \, .
%\end{align}

\subsection{Sandwiched state in a CFT} \label{sec-setup}

In this paper, we are interested in studying the refined R\'enyi relative entropy between states of a conformal field theory (CFT) on $R^{1,d-1}$ though our results are valid for CFT's on $R \times S^{d-1}$ as well. In particular, our approach will be to use the relation between the refined R\'enyi relative entropy and the relative entropy given in Eq.~\eqref{eq-ren-rel-fin}. To use this relation, we first need to construct the sandwiched state, $\r \,$, for given states $\rho$ and $\sigma$. In this section, we show that for a family of states $\rho$ and $\sigma$, the sandwiched state $\r$ can be prepared (up to a unitary transformation) by a Euclidean path integral over $\M' \, = \, S^{1} \times \hp$. 
%In Sec.~(\ref{sec-pert-analysis}), we will focus on a general CFT whereas in Sec.~(\ref{sec-non-pert-analysis}), we will specialize to CFT's with a semi-classical holographic dual. 
%\newline

Let us take the state $\sigma$ to be the vacuum state of the CFT, $\ket{\Omega} \, $, reduced to a spherical region, $B$, of radius $R$. The domain of dependence of $B$, which we denote by $\mathcal{D}(B) \, $, can be mapped to $\M \, = \, R \, \times \, \hp$ by a conformal transformation \cite{Casini:2011kv}. Under this transformation, the state $\sigma$ maps to a thermal state, $\tilde{\sigma} \, $, of temperature $T = 1/(2\pi R)$ on the hyperbolic space of radius $R$ \cite{Casini:2011kv}. Therefore, the state $\sigma$ is related to $\tilde{\sigma}$ by a unitary transformation. If we denote the Hamiltonian on the hyperbolic space by $H$, then the state $\sigma$ is given by
%\tcr{(Can one see already (without looking at the appendix) that it's supposed to be unitary?)}\tcb{(Yes. How else can density matrices can be related. If there is a map between two density matrices, then it has to be unitary. Otherwise, trace would not be conserved.)} \tcr{(Why can't it be something else, like a similarity transformation like $A^{-1} \rho A$ for some $A?$)}\tcb{(Is this Hermitian?)}.%
\begin{align}
U \, \sigma \, U^{\dagger} \, = \, \tilde{\sigma} \, \equiv \, \frac{e^{-2\pi R \, H}}{Z_{R}} \, , \label{eq-vac}
\end{align}
%\begin{align}
%    \sigma \, = \, \frac{U^{\dagger} \, e^{-2\pi R \, H} \, U}{Z_{R}} \, ,
%\end{align}
where $Z_{R} \, = \, \tr \, e^{-2\pi R \, H}$ is the thermal partition function on the hyperbolic space and $U$ is the unitary transformation that maps states from $B$ to hyperbolic space. 
%This implies that, up to a normalization constant, the modular Hamiltonian of $\sigma$, defined as $K_{\sigma} \, \equiv \, - \log \, \sigma \, $, is given by%related to the Hamiltonian of the CFT on a hyperbolic space. Using the conformal transformation from 
%\begin{align}
%    K_{\sigma} \, = \, 2\pi R \, U^{\dagger} H U \, . \label{eq-k-h}
%\end{align}
%Using the conformal transformation from $\M$ to $\mathcal{D}(B)$ and using the transformation law of stress-energy tensor, the modular Hamiltonian can be written as [cite]
%\begin{align}
%    K_{\sigma} \, = \, \int_{\mathcal{S}} \, d\Sigma_{B}^{b} \, \hat{\eta}^{a} \, T_{ab} \, , \label{eq-K-cov}
%\end{align}
%where we perform the integral over any achronal slice $\mathcal{S}$ within the domain of dependence of the spherical region $B$, and the vector $\hat{\eta}$ is the image of the time-translation vector \tcr{(Need to clarify what this is when this is rewritten.)} under the conformal transformation from $\M$ to $\mathcal{D}(B)$. In this paper, we will make use of both Eq.~\eqref{eq-k-h} and Eq.~\eqref{eq-K-cov} as required. \tcb{(Rewrite this later after finishing the non-perturbative section.)}
%This implies that the modular Hamiltonian generates a modular flow along the vector field $\hat{\eta}^{a}$ which remains inside the domain of independence of $B$.
%\newline

Following \cite{Lashkari-sand}, We now consider a state $\ket{\Psi} \, \equiv \, \mathcal{N} \, \Psi \ket{\Omega}$, where $\Psi$ is defined by the {smearing} of a local operator $\Psi(x)$ in a small neighborhood around the point $x$ and $\mathcal{N}$ is the normalization constant. We take state $\rho$ to be the state $\ket{\Psi}$ reduced to region $B$. The state $\rho$ can be represented by a Euclidean path integral on $R^{d}$ with open cuts just above and below the region $B$ and with insertions of operators $\Psi$ in the lower and in the upper half plane \footnote{For simplicity, we take $\Psi$ to be a Hermitian operator.}. 
%\newline

Just like the state $\sigma$, the state $\rho$ is also related to the state $\tilde{\rho}$  on the hyperbolic space by a unitary transformation. To find the state $\tilde{\rho} \, $, we use the conformal transformation from $R^{d}$ to $\M' \, = \, S^{1} \, \times \, \hp$ where the radii of $S^{1}$ and $\hp$ are equal to $R$. Note that this is simply the Euclidean version of the conformal transformation that maps $\mathcal{D}(B)$ to $\M$. Under this conformal transformation, the aforementioned path integral representation of the state $\rho$ maps to a path integral representation of a density matrix on $\M'$. We perform this analysis in Appendix~(\ref{app-rho}) and find that the matrix elements of $\rho$ and $\tilde{\rho}$, up to a normalization constant, can be written as
\begin{align}
\langle\tilde{\phi}_{-}| \, U \, \rho \, U^{\dagger} |\tilde{\phi}_{+}\rangle \, = \, \langle\tilde{\phi}_{-}| \, \tilde{\rho} \, |\tilde{\phi}_{+}\rangle \, \sim& \,  \int^{\Phi(\tau=2\pi R)=\tilde{\phi}_{-}}_{\Phi(\tau=0)=\tilde{\phi}_{+}} \, D\Phi \, e^{-I_{\text{0}}[\Phi]} \,\,\, \tilde{\Psi}(\tau = \pi R + \tau_{0}) \, \tilde{\Psi}(\tau = \pi R - \tau_{0}) \, ,  \label{eq-rho-cft-pi}
\end{align}
%only present the final result for the the path integral  states $\rho$ and $\tilde{\rho}$, up to a normalization constant, here:
for $0 \le \tau_{0} \le \pi R \, $. In this equation, $\tilde{\Psi} \, = \, U \, \Psi \, U^{\dagger}$ is the image of $\Psi$ under conformal transformation and $\tilde{\Psi}(\tau_{*})$ is the smearing of the local operator $\Psi(\tau,y)$ around a small neighborhood of $\tau = \tau_{*}$ and $y_{i} = 0$ where $0 < \tau <2\pi R$ and $y_{i}$ are the coordinates on $S^{1}$ and $\hp$ respectively. A pictorial representation of the path integral in Eq.~\eqref{eq-rho-cft-pi} is presented in Fig.~(\ref{fig-pi}). As we discuss in Appendix~(\ref{app-rho}), the path integral in Eq.~\eqref{eq-rho-cft-pi} can be written in operator language. This yields the following expression for states $\rho$ and $\tilde{\rho}$ 
\begin{align}
U \, \rho \, U^{\dagger} \, = \, \tilde{\rho} \, = \, \frac{\tilde{\sigma} \,\, \tilde{\Psi}(\pi R + \tau_{0}) \, \tilde{\Psi}(\pi R - \tau_{0})}{\langle \tilde{\Psi}(\pi R + \tau_{0}) \, \tilde{\Psi}(\pi R - \tau_{0}) \rangle_{\text{{$\scriptstyle \M'$}}}} \, , \label{eq-rho-cft}
\end{align}
where $\langle ... \rangle_{\text{{$\scriptstyle \M'$}}}$ is the correlation functions on $\M'$.
%\newline

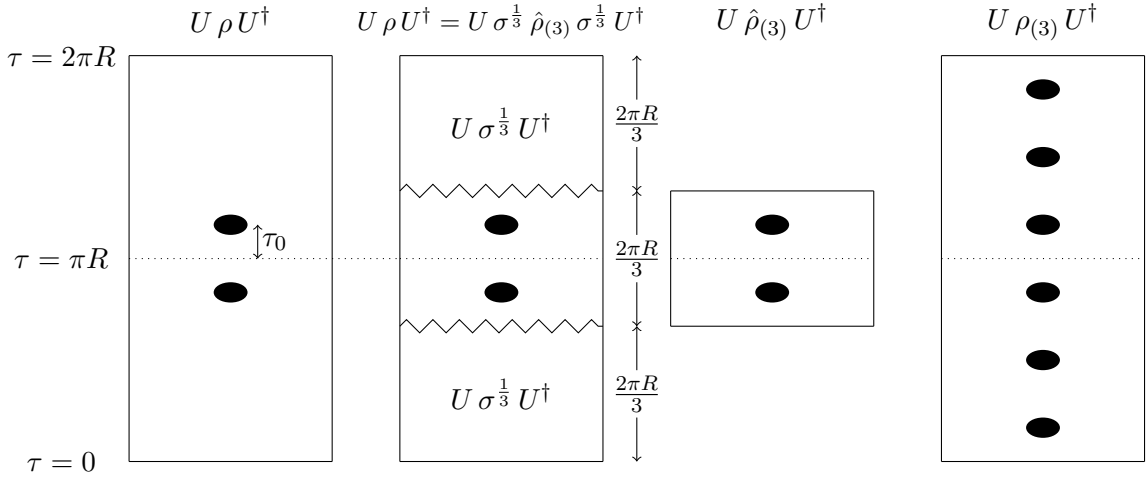
\begin{figure}
	\centering
	\begin{tikzpicture}[scale=0.9]
	\draw [black](-5,0) --(-2,0);
	\draw [black](-5,0) --(-5,6);
	\draw [black](-5,6) --(-2,6);
	\draw [black](-2,6) --(-2,0);
	\draw [<->](-3.1,3) --(-3.1,3.5);
	\node at (-2.85,3.25) {{$\tiny \tau_{0}$}};
	\fill[black] (-3.5,2.5) ellipse (.25cm and .15cm);%\node at (-2.5,2.5) {$A$};
	\fill[black] (-3.5,3.5) ellipse (.25cm and .15cm);%\node at (-2.5,3.5) {$A$};
	\node at (-3.5,6.5) {$U \, \rho \, U^{\dagger}$};
	\draw [dotted](-5,3) --(-1,3);
	%%%%%%%
	\draw [black](-1,0) --(2,0);
	\draw [black](-1,0) --(-1,6);
	\draw [black](-1,6) --(2,6);
	\draw [black](2,6) --(2,0);
	\fill[black] (0.5,2.5) ellipse (.25cm and .15cm);%\node at (-2.5,2.5) {$A$};
	\fill[black] (0.5,3.5) ellipse (.25cm and .15cm);
	%	\node at (1.5,2.5) {$A$};
	%	\node at (1.5,3.5) {$A$};
	\node at (0.5,5) {$U \, \sigma^{\frac{1}{3}} \, U^{\dagger}$};
	\node at (0.5,1) {$U \, \sigma^{\frac{1}{3}} \, U^{\dagger}$};
	\node at (0.5,6.5) {{{\small $U \, \rho \, U^{\dagger} = U \, \sigma^{\frac{1}{3}} \, \hat{\rho}_{(3)} \, \sigma^{\frac{1}{3}} \, U^{\dagger}$}}};
	\draw [decorate,decoration=zigzag](-1,4) --(2,4);
	\draw [decorate,decoration=zigzag](-1,2) --(2,2.0);
	\draw [dotted](-1,3) --(2,3);
	%%%%%%%%%%%%%%%%%%%
	\draw [black](3,2) --(6,2);
	\draw [black](3,2) --(3,4);
	\draw [black](3,4) --(6,4);
	\draw [black](6,4) --(6,2);
	\fill[black] (4.5,2.5) ellipse (.25cm and .15cm);%\node at (-2.5,2.5) {$A$};
	\fill[black] (4.5,3.5) ellipse (.25cm and .15cm);
	%	\node at (5.5,2.5) {$A$};
	%	\node at (5.5,3.5) {$A$};
	\node at (4.5,6.5) {$U \, \hat{\rho}_{(3)} \, U^{\dagger}$};
	\draw [dotted](3,3) --(6,3);
	\node at (2.5,3) {$\frac{2\pi R}{3}$};
	\draw [->] (2.5,3.35)--(2.5,4);
	\draw [->] (2.5,2.65)--(2.5,2);
	\node at (2.5,1) {$\frac{2\pi R}{3}$};
	\draw [->] (2.5,1.35)--(2.5,2);
	\draw [->] (2.5,0.65)--(2.5,0);
	\node at (2.5,5) {$\frac{2\pi R}{3}$};
	\draw [->] (2.5,5.35)--(2.5,6);
	\draw [->] (2.5,4.65)--(2.5,4);
	%%%%%%%%%%%%%%%%%%%
	\draw [black](7,0) --(10,0);
	\draw [black](7,6) --(7,0);
	\draw [black](7,6) --(10,6);
	\draw [black](10,6) --(10,0);
	\fill[black] (8.5,2.5) ellipse (.25cm and .15cm);%\node at (-2.5,2.5) {$A$};
	\fill[black] (8.5,3.5) ellipse (.25cm and .15cm);
	\fill[black] (8.5,4.5) ellipse (.25cm and .15cm);%\node at (-2.5,2.5) {$A$};
	\fill[black] (8.5,5.5) ellipse (.25cm and .15cm);
	\fill[black] (8.5,0.5) ellipse (.25cm and .15cm);%\node at (-2.5,2.5) {$A$};
	\fill[black] (8.5,1.5) ellipse (.25cm and .15cm);
	%\node at (9.5,2.5) {$A$};
	%\node at (9.5,3.5) {$A$};
	\node at (8.5,6.5) {$U \, \rho_{(3)} \, U^{\dagger}$};
	\draw [dotted](7,3) --(10,3);
	\fill[white] (10.5,2.5) ellipse (.25cm and .15cm);%\node at (-2.5,2.5) {$A$};
	%%%%%%%%%%%%%%%%%%
	\node at (-6,0) {$\tau = 0$};
	\node at (-6,3) {$\tau = \pi R $};
	\node at (-6,6) {$\tau = 2\pi R$};
	\end{tikzpicture}
	\caption{The pictorial representation of the construction of the states $\rho$ and $\r$ using a Euclidean path integral over $\M' \, = \, S^{1}\times \hp$. The vertical direction denotes the direction along the $S^{1}$ whereas the horizontal direction denotes the radial direction of  $\hp$ and we have suppressed the transverse $d-2$ dimensions. %(a) The state $U \, \rho \, U^{\dagger}$ can be prepared by a path integral with insertions of $2$ smeared operators; see Eq.~\eqref{eq-rho-cft-pi}. (b) The path integral of $U \, \rho \, U^{\dagger}$ in (a)  can be `cut' in the $S^{1}$ direction and can be written (for $n \ge 1$) as a matrix product $U \, \rho \, U^{\dagger} \, = \, U \, \sigma^{\frac{n-1}{2n}} \, \hat{\rho}_{(n)} \, \sigma^{\frac{n-1}{2n}} \, U^{\dagger}$. (c) This allows us to write $U \, \hat{\rho}_{(n)} \, U^{\dagger}$ as a path integral over $\M' = S^{1}\times \hp$, where the radius of $S^{1}$ is now $R/n$. (d) By gluing $n$ copies of $U \, \hat{\rho}_{(n)} \, U^{\dagger}$, we get a path integral representation of the sandwiched state, $U \, \r \, U^{\dagger}$, for $n \ge 1$. 
		(This figure is inspired by Fig.~($1$) of \cite{lashkari2016gravitational}.)}
	\label{fig-pi}
\end{figure}

Now consider the sandwiched state, $\r \,$, defined in Eq.~\eqref{eq-sand-st}.  %According to Eq.~\eqref{eq-ren-rel-fin}, the refined R\'enyi relative entropy between states $\rho$ and $\sigma$ is equal to the relative entropy between states $\r \,$ and $\sigma$. 
Note that we can write the sandwiched state, up to a normalization constant, as $\r \, \sim \, \hat{\rho}_{(n)}^{\, n} \,$, where
\begin{align}
\hat{\rho}_{(n)} \, = \, \sigma^{\frac{1-n}{2n}} \, \rho \, \sigma^{\frac{1-n}{2n}} \, .
\end{align}
Inverting this equation yields $\rho \, = \, \, \sigma^{\frac{n-1}{2n}} \, \hat{\rho}_{(n)} \, \sigma^{\frac{n-1}{2n}} \,$. As discussed in \cite{Lashkari:2014yva,Lashkari-sand,lashkari2016gravitational}, this expression for $\rho$ allows us to write a path integral representation of $\hat{\rho}_{(n)} \,$. This follows from the observation that for $1 \, < n < \frac{\pi R}{\tau_{0}}$ \footnote{We can follow \cite{lashkari2016gravitational} and take the limit $\tau_{0} \to 0$ or we can follow \cite{Ugajin:2018rwd} and take $R \to 0$. This will allow us to use the following construction for arbitrarily large $n$.}, the path integral representation of the matrix element in Eq.~\eqref{eq-rho-cft-pi} can be can be cut in $S^{1}$ direction as shown in Fig.~(\ref{fig-pi}). This yields that %the operator \tcr{(state?)}\tcb{(NO. It's not normalized.)} 
$\hat{\rho}_{(n)}$ can be written as a path integral (with two operator insertions) over $S^{1} \times \hp$ where the radius of $S^{1}$ is now $R/n$. Now since $\r \, \sim \hat{\rho}_{(n)}^{\, n} \, $, we take $n$ copies of $\hat{\rho}_{(n)}$ and glue them together. As a result, we find that, for integer $n \ge 1$, the sandwiched state $\r \,$ can be written as a path integral over $\M'$ with $2n$ operators insertions \cite{Lashkari:2014yva,Lashkari-sand,lashkari2016gravitational}, as shown in Fig.~(\ref{fig-pi}). %Note that the sandwiched state, $\r$, has a $Z_{n}$ symmetry as is evident in Fig.~(\ref{fig-pi}).
%\newline

In this paper, one of our goals is to use Eq.~\eqref{eq-ren-rel-pert-fin} to study the refined R\'enyi relative entropy when the state $\rho$ is perturbatively expanded around the state $\sigma$ as in Eq.~\eqref{eq-rho-ser}. To write the state $\rho$ as in Eq.~\eqref{eq-rho-ser}, we take the operator $\Psi$ as a perturbation  around the identity and write it as
\begin{align}
\Psi \, = \, 1 \, + \, \epsilon \, \OO \, + \, O(\epsilon^{2}) \, .
\end{align}
Inserting this in Eq.~\eqref{eq-rho-cft} and comparing it with Eq.~\eqref{eq-rho-ser}, we find
\begin{align}
U \, \rho^{(1)} \, U^{\dagger} \, = \, \tilde{\rho}^{(1)} \, = \, \tilde{\sigma} \,\, \tilde{\OO}(\pi R + \tau_{0}) \, + \, \tilde{\sigma} \,\, \tilde{\OO}(\pi R - \tau_{0}) \, . \label{eq-rho-one-cft}
\end{align}
Since $\tilde{\OO}$ are smeared operators, we can equivalently write Eq.~\eqref{eq-rho-one-cft} as an integral over $\M'$,
\begin{align}
U \, \rho^{(1)} \, U^{\dagger} \, = \, \tilde{\rho}^{(1)} \, = \, \int_{0}^{2\pi R} \, d\tau \, \int_{H^{d-1}} \, d^{d-1}y \,\,\, \tilde{\lambda}(\tau,y) \,\, \tilde{\sigma} \,\, \tilde{\OO}(\tau, y) \, , \label{eq-rho-one-cft-fin}
\end{align}
where $\tilde{\lambda}$ is equal to the smearing function inside small neighborhoods of $\tau \, = \, \pi R \pm \tau_{0}$ and $y_{i}=0$ and vanishes everywhere outside these two neighborhoods. Also note that, by construction, the function $\tilde{\lambda}$ is symmetric around $\tau \, = \, \pi R \, $.
%\newline

This finishes our discussion of the path integral construction of sandwiched state, $\r$. In the next two sections, we will study the refined R\'enyi relative entropy between these state $\rho$ given in Eq.~\eqref{eq-rho-cft} and state $\sigma$ in Eq.~\eqref{eq-vac}. %$\sigma$ and $\rho$. We will study the refined R\'enyi relative entropy between these two states in the next two sections. \tcb{(This is no longer valid. So rewrite this line.)}

\section{Perturbative refined R\'enyi relative entropy in a general CFT} \label{sec-pert-analysis}

The perturbative calculation of relative entropy between the vacuum state and a slightly perturbed state of a CFT, both reduced to a spherical region, has been well-studied \cite{Faulk-GR-entanglement,Faulkner:2014jva,Lashkari:2015hha}. It was shown in \cite{Faulk-GR-entanglement} that the relative entropy at the lowest order in the perturbation parameter can be written as the symplectic flux of a scalar field through a Cauchy slice of an \textit{auxiliary} AdS spacetime, as given in Eq.~\eqref{xx}. Note that this result was derived purely from CFT calculations without assuming the AdS/CFT correspondence. In fact, this result is true for \textit{all} CFT's. In this sense, this result is a generalization of \cite{Lashkari:2015hha} where the same result was derived for holographic CFT's using the AdS/CFT correspondence. 
%\newline

Our goal in this section is to study the refined R\'enyi relative entropy between states of a general CFT reduced to a spherical region $B$ and, for simplicity, we take the radius of $B$ to be $R \, = \, 1$. We take state $\sigma$ to be the reduced vacuum state as in Eq.~\eqref{eq-vac} and $\rho$ to be the perturbation around $\sigma$ as in Eq.~\eqref{eq-rho-ser} and Eq.~\eqref{eq-rho-one-cft-fin}. %We use the general perturbative formula for the RRRE, Eq.~\eqref{eq-ren-rel-pert-fin}, to study the 
In the following, we show that refined R\'enyi relative entropy at lowest order in the perturbation parameter can also be written as the symplectic flux of some scalar field through a Cauchy slice of the aforementioned auxiliary AdS-spacetime. Hence, our result can be considered to be a one-parameter generalization of the result of \cite{Faulk-GR-entanglement} given in  Eq.~\eqref{xx}. 
%\newline

As we discussed in Sec.~(\ref{sec-setup}), the states $\sigma$ and $\rho$ are related to the states on the hyperbolic space by a unitary transformation. Using the invariance of the refined R\'enyi relative entropy under unitary transformations, we write
\begin{align}
\S_{n} (\rho||\sigma) \, = \, \S_{n} (U^{\dagger} \, \tilde{\rho} \, U ||U^{\dagger} \, \tilde{\sigma} \, U) \, = \, \S_{n} (\tilde{\rho}||\tilde{\sigma}) \, ,
\end{align}
where $\tilde{\sigma}$ and $\tilde{\rho}$ are states on the hyperbolic space and their explicit forms are given in Eq.~\eqref{eq-vac} and Eq.~\eqref{eq-rho-cft} respectively.
%\newline

\begin{figure}
	\centering
	\includegraphics[scale=1.25]{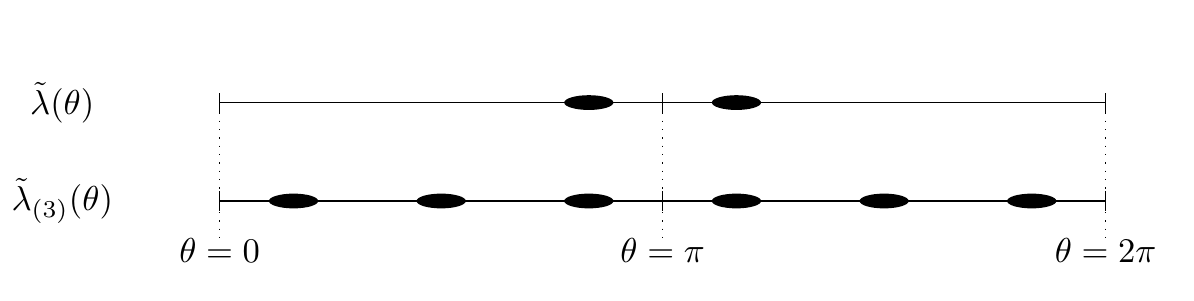}
	\caption{A pictorial representation of the source function $\tilde{\lambda}(\tau,y)$ and the $\Z_{3}$-symmetric source function $\tilde{\lambda}_{(3)}(\tau,y)$ where we have suppressed the $y$-coordinates for clarity. The functions are non-zero everywhere except in the shaded regions. Both the source function $\tilde{\lambda}$ and the $\Z_{3}$-symmetric source function $\tilde{\lambda}_{(3)}$ are symmetric around $\tau = \pi$. Whereas the source function $\tilde{\lambda}$ is non-zero only in the neighborhood of two points, the $\Z_{3}$-symmetric source function $\tilde{\lambda}_{(3)}$ is non-zero in the neighborhood of six points and has a periodicity of $2\pi/3 \, $.  }
	\label{fig-lambda-n}
\end{figure}

Now to compute the perturbative refined R\'enyi relative entropy, we use the general formula derived in  Eq.~\eqref{eq-ren-rel-pert-fin}.  By inserting the perturbation of $\tilde{\rho}$ given in Eq.~\eqref{eq-rho-one-cft-fin} in the general formula, we get \footnote{Note that we have taken $R=1$ here.}
\begin{align}
\S^{(2)}_{n} (\rho||\sigma) \, = \, - \, \,&\sum_{k=0}^{n-1}\sum_{j=0}^{n-1} \,\, \left( \prod_{i=1}^{2} \int_{0}^{2\pi} \, d\tau_{i} \, \int_{H^{d-1}} \, d^{d-1}y_{i} \,\,\, \tilde{\lambda}(\tau_{i},y_{i})\right)   \,\, \int_{-\infty}^{\infty} \, \frac{ds}{4 \,\sinh^{2}\left(\frac{s+i\delta}{2}\right)} \nonumber\\  \,\,\, & \times \,\, \tr \Big( \ts^{-\frac{is}{2\pi}+\frac{(k-j)}{n}} \, \tO(\tau_{1},y_{1}) \, \ts^{1+\frac{is}{2\pi}-\frac{(k-j)}{n}} \, \tO(\tau_{2},y_{2}) \Big) \, , \label{eq-sn-cft-pert-int}
%\tr \Big( \sigma^{-\frac{is}{2\pi}+\frac{(k-j)}{n}} \, \OO(\theta_{1},y_{1}) \, \sigma^{1+\frac{is}{2\pi}-\frac{(k-j)}{n}} \, \OO(\theta_{2},y_{2}) \Big) \, , \label{eq-sn-cft-pert-int}
\end{align}
where $\S^{(2)}_{n} (\rho||\sigma)$ is the refined R\'enyi relative entropy at the second order in the perturbation parameter as defined in Eq.~\eqref{eq-sn-ser-exp}. Let us now try to simplify the above expression. We start by using Eq.~\eqref{eq-vac} and the Euclidean evolution
\begin{align}
\ts^{a} \, \tO(\tau,y) \, \ts^{-a} \, = \, e^{-2\pi a H} \, \tO(\tau,y) \, e^{2\pi a H} \, = \, \tO(\tau - 2\pi a ,y) \, ,
\end{align}
to write Eq.~\eqref{eq-sn-cft-pert-int} as
\begin{align}
\S^{(2)}_{n} (\rho||\sigma) \, = \, - \, \,&\sum_{k=0}^{n-1}\sum_{j=0}^{n-1} \,\, \left( \prod_{i=1}^{2} \int_{0}^{2\pi} \, d\tau_{i} \, \int_{H^{d-1}} \, d^{d-1}y_{i} \,\,\, \tilde{\lambda}(\tau_{i},y_{i})\right)   \,\, \int_{-\infty}^{\infty} \, \frac{ds}{4 \,\sinh^{2}\left(\frac{s+i\delta}{2}\right)} \nonumber\\  \,\,\, & \times \,\, \tr \Big( \ts  \, \tO(\tau_{2} + 2\pi k/n , y_{2}) \,  \tO(\tau_{1} + 2\pi j/n + i s ,y_{1}) \Big) \, . \label{eq-sn-cft-pert-int-2}
\end{align}
Now we write the above equation in terms of  a time-ordered correlation function on $\M' \, = \, S^{1} \times \hp \, $ as \cite{Faulk-GR-entanglement}
\begin{align}
\S^{(2)}_{n} (\rho||\sigma) \, = \, - \, \,&\sum_{k=0}^{n-1}\sum_{j=0}^{n-1} \,\, \left( \prod_{i=1}^{2} \int_{0}^{2\pi} \, d\tau_{i} \, \int_{H^{d-1}} \, d^{d-1}y_{i} \,\,\, \tilde{\lambda}(\tau_{i},y_{i})\right)   \,\, \int_{-\infty}^{\infty} \, \frac{ds}{4 \,\sinh^{2}\left(\frac{s+i\delta \, \text{sgn}[\tau_{2k}-\tau_{1j}]}{2}\right)} \nonumber\\  \,\,\, & \times \,\,  \Big\langle \mathcal{T}  \,\, \tO(\tau_{2} + 2\pi k/n , y_{2}) \,  \tO(\tau_{1} + 2\pi j/n + i s ,y_{1}) \Big\rangle_{\text{{$\scriptstyle \M'$}}} \, , \label{eq-sn-cft-pert-int-3}
\end{align}
where $\tau_{2k} \, \equiv \, (\tau_{2} + 2\pi k/n) \, \text{mod } 2\pi \, $ and $\tau_{1j}$ is defined similarly. Now by using the KMS condition \cite{k,ms,Haag1967} on the periodicity of the thermal two-point function 
\begin{align}
\Big\langle \mathcal{T}  \,\, \tO(\tau_{2} + 2\pi , y_{2} ) \,  \tO(\tau_{1}  ,y_{1}) \Big\rangle_{\text{{$\scriptstyle \M'$}}} \, = \, \Big\langle \mathcal{T}  \,\, \tO(\tau_{2} , y_{2} ) \,  \tO(\tau_{1}  ,y_{1}) \Big\rangle_{\text{{$\scriptstyle \M'$}}} \, ,
\end{align}
and by redefining the angular coordinates, we simplify Eq.~\eqref{eq-sn-cft-pert-int-3} to get  %\tcr{(I guess it doesn't really clarify what the condition is, who KMS are and if there is supposed to be a citation.} {\tcb{ Is not KMS condition a well known thing? Do we really need to write it explicitly? )} 
\begin{align}
\S^{(2)}_{n} (\rho||\sigma) \, = \, - \, \, &\left( \prod_{i=1}^{2} \int_{0}^{2\pi} \, d\tau_{i} \, \int_{H^{d-1}} \, d^{d-1}y_{i} \,\,\, \tilde{\lambda}_{(n)}(\tau_{i},y_{i})\right)   \,\, \int_{-\infty}^{\infty} \, \frac{ds}{4 \,\sinh^{2}\left(\frac{s+i\delta \, \text{sgn}[\tau_{2}-\tau_{1}]}{2}\right)} \nonumber\\  \,\,\, & \times \,\,  \Big\langle \mathcal{T}  \,\, \tO(\tau_{2} , y_{2}) \,  \tO(\tau_{1} + i s ,y_{1}) \Big\rangle_{\text{{$\scriptstyle \M'$}}} \,\, , \label{eq-sn-cft-pert-fin}
\end{align}
where we have absorbed all the $n$ dependence in a $\Z_{n}$-symmetric source function, which we define as\footnote{The function $\tilde{\lambda}(\tau,y)$ in Eq.~\eqref{eq-rho-one-cft-fin} was defined only in the domain $\tau = [0,2\pi)$. We extend this function beyond this domain by demanding the periodicity condition: $\tilde{\lambda}(\tau+2\pi , y) \, = \, \tilde{\lambda}(\tau , y)$.}
\begin{align}
\tilde{\lambda}_{(n)}(\tau , y) \, \equiv \, \sum_{k=0}^{n-1} \, \tilde{\lambda}(\tau - 2\pi k/n , y ) \, . \label{eq-zn-source}
\end{align}
Recall from Sec.~(\ref{sec-setup}) that the source function, $\tilde{\lambda}$, was only non-zero in the small neighborhood of two points. On the contrary, the $\Z_{n}$-symmetric source function, $\tilde{\lambda}_{(n)}$, has a periodicity of $2\pi/n$ and is non-zero in the neighborhood of $2n$ points. Just like the source function, $\tilde{\lambda}$, the $\Z_{n}$-symmetric source function, $\tilde{\lambda}_{n}$, is also symmetric around $\tau \, = \, \pi$. A pictorial representation of the $\Z_{n}$-symmetric source function, $\tilde{\lambda}_{n}$, is presented in Fig.~(\ref{fig-lambda-n}).  %\tcb{(COMPARE WITH THE PATH INTEGRAL REPRESENTATION OF THE SANDWICHED STATE IN FIG.~\ref{fig-pi}) }
%\newline

In the following, we use Eq.~\eqref{eq-sn-cft-pert-fin} to write the perturbative refined R\'enyi relative entropy as the symplectic flux. Note that an expression similar to Eq.~\eqref{eq-sn-cft-pert-fin} was found in \cite{Faulk-GR-entanglement} for the perturbative relative entropy. In fact, Eq.~\eqref{eq-sn-cft-pert-fin} reduces to the expression of perturbative relative entropy in \cite{Faulk-GR-entanglement} if we replace $\tilde{\lambda}_{(n)}$ with $\tilde{\lambda}$. \ This should not be surprising given that the refined R\'enyi relative entropy approaches the relative entropy in the limit $n \to 1$ and all the $n$ dependence in Eq.~\eqref{eq-sn-cft-pert-fin} is encoded in the $\Z_{n}$-symmetric source function, $\tilde{\lambda}_{(n)} \, $. This is a useful observation as it implies that we can write perturbative refined R\'enyi relative entropy in Eq.~\eqref{eq-sn-cft-pert-fin} in terms of the symplectic flux by simply changing the boundary conditions of the scalar field in Eq.~\eqref{xx} from $\tilde{\lambda}$ to $\tilde{\lambda}_{(n)}$.  %replacing $\tilde{\lambda}$ with $\tilde{\lambda}_{(n)}$ in Eq.~\eqref{xx}.
For the sake of completeness, we will now briefly sketch the argument of \cite{Faulk-GR-entanglement} and derive the expression of the perturbative refined R\'enyi relative entropy in terms of the symplectic flux in the following. 

\subsection{Refined R\'enyi relative entropy as a symplectic flux} \label{sec-rrre-sf}

Let us consider a scalar in a fixed AdS-Rindler wedge. Recall that the metric of the AdS-Rindler wedge can be written as
\begin{align}
ds^{2} \, = \, - (r^{2}-1) \, dt^{2} + \frac{dr^{2}}{r^{2}-1} + r^{2} \, ds^{2}_{\hp} \, . \label{eq-ads-rind}
\end{align}
For a CFT with a holographic dual, the AdS-Rindler wedge is precisely the bulk region that is dual to the vacuum state reduced to a spherical region. In this section, however, we treat it as an auxiliary spacetime without assuming the AdS/CFT correspondence.
%\newline

Note that, after a Wick rotation, the AdS-Rindler patch in Eq.~\eqref{eq-ads-rind} becomes a Euclidean black hole with $\M' = S^{1} \times \hp$ as the asymptotic boundary. The real time solution of the Lorentzian scalar field equation can be written in terms of a bulk-to-boundary Euclidean propagator as \cite{Witten:1998qj,Marolf:2017kvq} 
\begin{align}
\phi(r,t,y) \, = \, \int_{0}^{2\pi } \, d\tau' \, \int_{\hp} d^{d-1}y' \,\,\, \phi_{0}(\tau',y') \,\, K_{E}\big(r,it,y \, | \, \tau',y' \, \big) \, , \label{eq-ke}
\end{align}
where $ \{ \tau , y \}$ are the coordinates on $\M'$ and the bulk point in the bulk-to-boundary Euclidean propagator, $K_{E}$, is analytically continued to real time. The asymptotic boundary condition at $r \to \infty$ of the field, after analytic continuation, is fixed  by $\phi_{0}(\tau,y)$.  
%\newline

The main insight that was used in \cite{Faulk-GR-entanglement} to relate the perturbative relative entropy to symplectic flux was that the two point function on $\M'$ can be written as the symplectic flux of the scalar field on a constant radial slice ($r =r_{0}$) of the AdS-Rindler wedge. That is, \cite{Faulk-GR-entanglement}
\begin{align}
\Big\langle \mathcal{T} \,\,  \tO(\tau,y_{1}) \,  \tO(i s,y_{2}) \,  \Big\rangle_{\text{{$\scriptstyle \M'$}}} \, = \, - \, \int_{-\infty}^{\infty} dt \, \int_{\hp} d^{d-1}y \,\,\, \omega_{\phi}\Big( \, K_{E}\big(r_{0},it,y \, | \, \tau,y_{1} \, \big) \, , \, K_{R}\big(r_{0} , t , y \, | \, s , y_{2} \, \big) \,  \Big) \, , \label{eq-corr-symp}
\end{align}
where $K_{R}$ is the retarded bulk-to-boundary propagator which is related to the Euclidean bulk-to-boundary propagator according to
\begin{align}
K_{R}\big(r , t , y \, | \, s , y' \, \big) \, = \, i \Theta(t-s) \, \lim_{\mu \to 0^{+}} \, \left[  K_{E}\big(r , it , y \, | \, i s -\mu  , y' \, \big) -  K_{E}\big(r , it , y \, | \, i s + \mu , y' \, \big) \right] \, .
\end{align}
%\newline

Now we insert Eq.~\eqref{eq-corr-symp} in Eq.~\eqref{eq-sn-cft-pert-fin} and write the perturbative refined R\'enyi relative entropy as
\begin{align}
\S^{(2)}_{n} (\rho||\sigma) \, = \, \, &\left( \prod_{i=1}^{2} \int_{0}^{2\pi} \, d\tau_{i} \, \int_{H^{d-1}} \, d^{d-1}y_{i} \,\,\, \tilde{\lambda}_{(n)}(\tau_{i},y_{i})\right)   \,\, \int_{-\infty}^{\infty} \, \frac{ds}{4 \,\sinh^{2}\left(\frac{s+i\delta \, \text{sgn}[\tau_{2}-\tau_{1}]}{2}\right)} \nonumber\\  \,\,\, & \times \,\,  \int_{-\infty}^{\infty} dt \, \int_{\hp} d^{d-1}y \,\,\, \omega_{\phi}\Big( \, K_{E}\big(r_{0},it,y \, | \, \tau_{2} - \tau_{1} ,y_{2} \, \big) \, , \, K_{R}\big(r_{0} , t , y \, | \, s , y_{1} \, \big) \,  \Big) \,\, . \label{eq-sn-cft-symp}
\end{align}
The integral over $s$ was computed in \cite{Faulk-GR-entanglement}. We use that result here and write the perturbed refined R\'enyi relative entropy as
\begin{align}
\S^{(2)}_{n} (\rho||\sigma) \, = \, -2\pi \, &\left( \prod_{i=1}^{2} \int_{\M'} \, dX_{i} \,\,\, \tilde{\lambda}_{(n)}(X_{i})\right) \,\,\, \int_{\Sigma_{0}} \,\,\, \omega_{\phi}\Big( \, K_{E}\big(r,0,y \, | \, X_{2} \, \big) \, , \, \partial_{t} K_{E}\big(r , 0 , y \, | \, X_{1} \, \big) \,  \Big) \,\, , \label{eq-sn-cft-symp-2}
\end{align}
where $\Sigma_{0}$ is the $t=0$ slice in the AdS-Rindler patch and where $X \, = \, \{\tau,y\}$ are the coordinates on $\M'$. Since the symplectic current $\omega_{\phi}$ is bilinear, we write Eq.~\eqref{eq-sn-cft-symp-2} as
\begin{align}
\S^{(2)}_{n} (\rho||\sigma) \, = \,\, \int_{\Sigma_{0}} \,\,\, \omega_{\phi}\Big( \, \Phi\big(r , t , y \big) \, , \, \lie_{\xi} \Phi\big(r , t , y \big) \,  \Big) \,\, , \label{eq-sn-cft-symp-fin}
\end{align}
where $\xi^{a} \, = \, -2\pi (\partial_{t})^{a}$ is a killing vector field of the AdS-Rindler wedge as in Eq.~\eqref{eq-eta-intro-2} and where we have defined %is normalized such that $\partial_{\eta} \, = \, -2\pi \partial_{t} \, $. Moreover, 
\begin{align}
\Phi\big(r , t , y \big) \, = \, \int_{\M'} \, dX \,\,\, \tilde{\lambda}_{(n)}(X) \,\, K_{E}\big(r,t,y \, | \, X \, \big) \, .
\end{align}
By comparing this with Eq.~\eqref{eq-ke}, we deduce that $\Phi(r,t,y)$ is the solution of the scalar field equation in bulk and the boundary condition of its analytic continuation is fixed by the $\Z_{n}$-symmetric source function, $\tilde{\lambda}_{(n)}$, defined in Eq.~\eqref{eq-zn-source}.
%\newline

Eq.~\eqref{eq-sn-cft-symp-fin} is the main result of this section. It shows that for integer $n\ge 1$, the perturbative refined R\'enyi relative entropy between the states $\rho$ in Eq.~\eqref{eq-rho-cft} and $\sigma$ in Eq.~\eqref{eq-vac} is related to the symplectic flux of a scalar field through a Cauchy slice of the AdS-Rindler spacetime. More interestingly, it shows that all the dependence in the R\'enyi parameter $n$ is encoded in the $\Z_{n}$-symmetric source function, $\tilde{\lambda}_{n} \, $, which provides the boundary conditions for the scalar field. Also note that we recover Eq.~\eqref{xx} from Eq.~\eqref{eq-sn-cft-symp-fin} by taking the limit $n \to 1$. Hence, our result is a one-parameter generalization of Eq.~\eqref{xx}.

%The relation between the symplectic flux of a scalar field and the perturbed RRRE given in Eq.~\eqref{eq-sn-cft-symp-fin} is the main result of this section. To summarize, we \tcb{state (better word?)} that the RRRE between the states $\rho$ in Eq.~\eqref{eq-rho-cft} and $\sigma$ in Eq.~\eqref{eq-vac} is given by the symplectic flux of a scalar field through a Cauchy slice of the AdS-Rindler spacetime and that all the dependence in the R\'enyi parameter $n$ is encoded in the boundary condition of the scalar field. 

\section{Refined R\'enyi relative entropy in a holographic CFT} \label{sec-non-pert-analysis}

We now focus on holographic CFT's on $R^{1,d-1}$. The relative entropy between an arbitrary state and a vacuum state reduced to a spherical region was studied in \cite{lashkari2016gravitational}, where it was shown that the holographic dual of the relative entropy between these states is equal to the difference between certain conserved charges in the bulk dual of these states, as stated in Eq.~\eqref{yy}. %\tcb{(rewrite this later. Review their results in the introduction.)} %In this section, we derive a similar holographic relation for the RRRE. More precisely, we 
%\newline

In this section, we are interested in the refined R\'enyi relative entropy between two states of a holographic CFT reduced to a spherical region $B$. In particular, we will study the refined R\'enyi relative entropy between the state $\rho \,$ in Eq.~\eqref{eq-rho-cft} and the vacuum state, $\sigma \,$, given in Eq.~\eqref{eq-vac}. Moreover, we take the radius of the region $B$ to be $R \, = \, 1$. We will show that the refined R\'enyi relative entropy between these states satisfies a holographic relation similar to Eq.~\eqref{yy}. Our result will be a one-parameter generalization of the result of \cite{lashkari2016gravitational} given in Eq.~\eqref{yy}. 
%\newline

Note that we can simply deduce the holographic formula for refined R\'enyi relative entropy using Eq.~\eqref{eq-ren-rel-fin} and Eq.~\eqref{yy}. According to Eq.~\eqref{eq-ren-rel-fin}, the refined R\'enyi relative entropy between $\rho$ and $\sigma$ is equal to the relative entropy between the sandwiched state, $\r \,$, and $\sigma$. If the sandwiched state, $\r \,$, has a semi-classical gravity dual and if there exits a vector field in the bulk spacetime that satisfies boundary conditions similar to those in Eq.~\eqref{eq-eta-bc-intro} and Eq.~\eqref{eq-binormal}, then Eq.~\eqref{yy} implies that the relative entropy between $\r$ and $\sigma$ is equal to difference of conserved charges in the bulk duals of $\r$ and $\sigma$. Then, by Eq.~\eqref{eq-ren-rel-fin}, this is precisely the holographic expression for the refined R\'enyi relative entropy between states $\rho$ and $\sigma$. 
%\newline

\begin{figure}
	\centering
	\begin{tikzpicture}
	\node at (-2,0) {\includegraphics[scale=0.65]{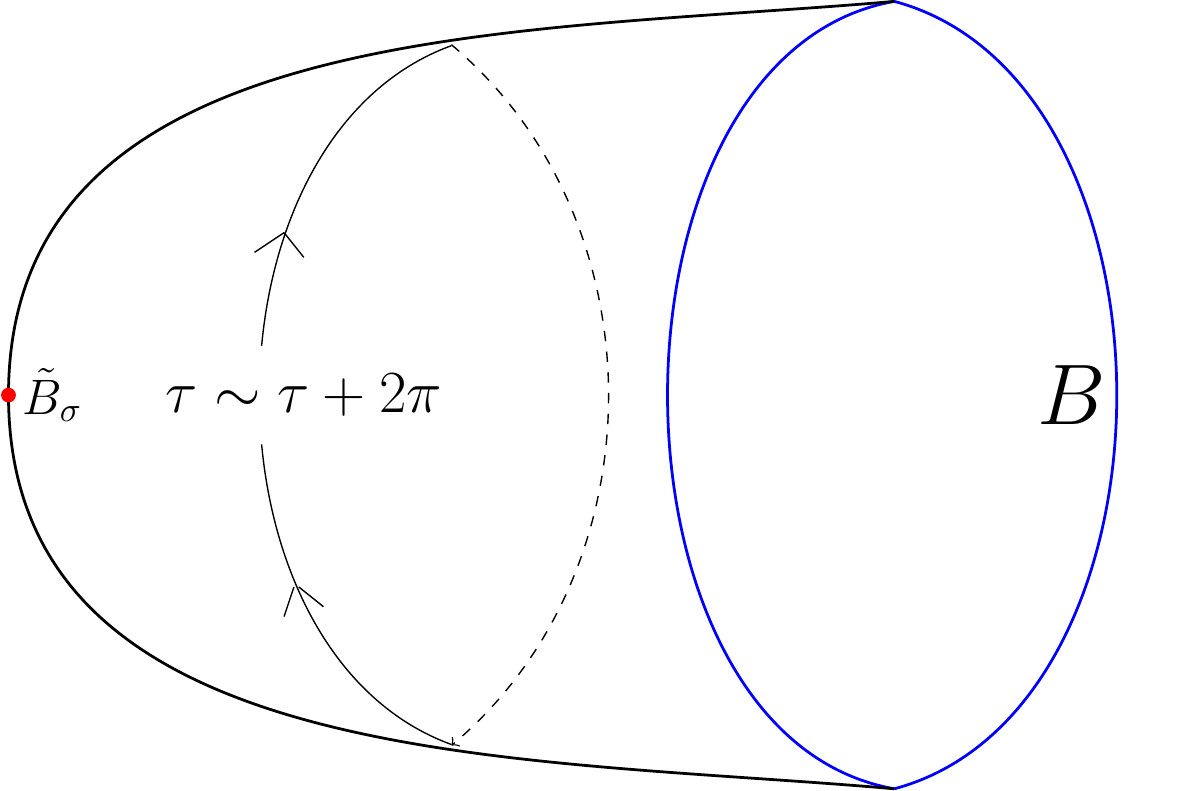}};
	\node at (7,0) {\includegraphics[scale=0.65]{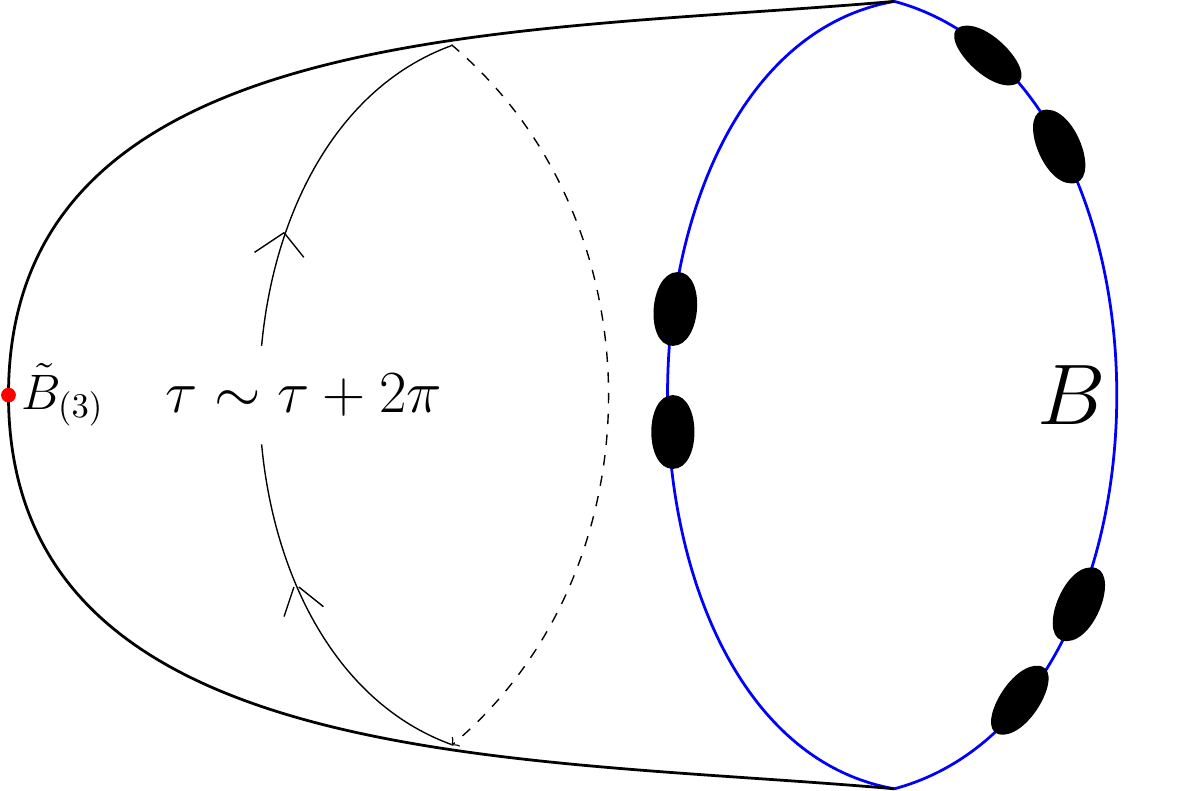}};
	\node at (-2.5,3.5) {\huge $\MM_{\sigma}$};
	\node at (6.5,3.5) {\huge $\MM_{(3)}$};
	\end{tikzpicture}
	\caption{The Euclidean bulk geometries dual to the boundary states $\sigma$ (\textit{left}) and $\rho_{(3)}$ (\textit{right}). The boundary $B$ (shown in blue) is $S^{1}\times \hp$. The Euclidean circle on the boundary contracts as one goes deep into the bulk and it is shrinks to zero size at a codimension-$2$ surface marked with a red dot. The shaded region on the boundary of $\MM_{(3)}$ denotes the smeared operators inserted in the path integral representation of the sandwiched state, $\rho_{(3)}$; see Fig.~(\ref{fig-pi}). The smearing function provides the $\Z_{3}$-symmetric boundary conditions for the bulk matter fields. (This figure is inspired by Fig.~($2$) of \cite{lashkari2016gravitational}.) }
	\label{fig-euc-geom}
\end{figure}

Motivated by the above observation, we now discuss the bulk dual of the  sandwiched state $\r$. %We start by reviewing the bulk dual of the vacuum state, Eq.~\eqref{eq-vac} \tcb{(Remove this later), and that of the state $\rho$ in Eq.~\eqref{eq-rho-cft}}. T
First, recall that the bulk region corresponding to the vacuum state reduced to a spherical region $B$ is the AdS-Rindler wedge \cite{Casini:2011kv}, which is described by the metric %\footnote{In general, AdS length scale is equal to the radius of the spherical boundary region $B$, which we have set to be $R \, = \, 1$.}  
given in Eq.~\eqref{eq-ads-rind}. 
%\begin{align}
%
%\end{align}
The codimension-$2$ bifurcation surface, given by $r \, = \, 1$ for any finite $t$ in the metric Eq.~\eqref{eq-ads-rind}, corresponds to the extremal surface  %Hubeny-Rangamany-Takayanagi (HRT) surface 
corresponding to the boundary region $B$. We denote this surface by $\tilde{B}_{\sigma}$. Under Wick rotation, the metric of the AdS-Rindler wedge in Eq.~\eqref{eq-ads-rind} becomes
\begin{align}
ds^{2} \, = \, (r^{2}-1) \, d\tau^{2} + \frac{dr^{2}}{r^{2}-1} + r^{2} \, ds^{2}_{\hp} \, , \label{eq-ads-rind-2}
\end{align}
where $0 \le \tau < 2\pi \,$ is imposed to ensure that there is no canonical singularity. This is the metric of the Euclidean black hole whose asymptotic boundary is $\M'\, = \, S^{1} \times \hp\,$. Note that $S^{1}$ contracts as we go deep into the bulk and it shrinks to zero size at a codimension-$2$ bifurcation surface, $\tilde{B}_{\sigma}$. This Euclidean geometry can be represented as a cigar geometry as shown in Fig.~(\ref{fig-euc-geom}). % [CITE]. %\tcb{(Include Figure.)}
%\newline

Now recall from Fig.~(\ref{fig-pi}) that for integer $n\ge 1$, the sandwiched state $\r$ can be prepared by a path integral over $\M'$ with $2n$ operator insertions. The Euclidean bulk dual of this state can be found by solving the Euclidean bulk equations of motions with appropriate boundary conditions. The boundary condition for the bulk Euclidean metric is that it approaches the metric of $\M'$ near the asymptotic boundary. We also impose the condition that the Euclidean time, $\tau$, has a periodicity of $2\pi$. Moreover, the sources for the operators inserted on the boundary provide the boundary conditions for the bulk matter fields according to the standard GKPW dictionary \cite{Gubser:1998bc,Witten:1998qj}. This Euclidean geometry is shown in Fig.~(\ref{fig-euc-geom}). 
%\newline

Note that just like in the Euclidean black hole in Eq.~\eqref{eq-ads-rind-2}, the Euclidean circle in the Euclidean bulk dual of the sandwiched state contracts as we go into the bulk. We denote the codimension-$2$ surface where the Euclidean circle shrinks to zero size by $\tilde{B}_{(n)}$. A general ansatz for the Euclidean metric near $\tilde{B}_{(n)}$ can be written as \cite{lashkari2016gravitational}
\begin{align}
ds^{2} \, = \, \alpha_{(n)}^{2}(\tau,\hat{r}) \, d\tau^{2} + d\hat{r}^{2} + 2 \beta^{(n)}_{i}(\tau,\hat{r}) \, d\tau dy^{i} + h_{ij} dy^{i} dy^{j} \, ,\label{eq-ads-sand}
\end{align}
with $\alpha_{(n)}(\tau,\hat{r}) \, = \, \hat{r} + O(\hat{r}^{2})$ and $\beta^{(n)}_{i}(\tau,\hat{r}) \, = \, O(\hat{r}^{2})\,$ , where $\hat{r}$ is the proper distance from $\tilde{B}_{(n)}$ and $y^{i}$ and $h_{ij}$ are coordinates and the induced metric on $\tilde{B}_{(n)} \,$. %This Euclidean geometry is shown in Fig.~(\ref{fig-euc-geom}). 
%Note that the vector field, $\tau^{a} \, \equiv \, (\partial_{\tau})^{a} \, $, which is a killing vector field of the Euclidean black hole in Eq.~\eqref{eq-ads-rind-2}, is not a killing vector field of the metric in Eq.~\eqref{eq-ads-sand}.
%\newline

Note that the Euclidean vector field $\xi_{E}^{a} \, = \, 2\pi \, (\partial_{\tau})^{a} \, $ vanishes at $\tilde{B}_{(n)}$ and it satisfies 
\begin{align}
\nabla^{a}\xi_{E}^{b} \, - \, \nabla^{b}\xi_{E}^{a} \, = \, 4\pi \, \Big( (\partial_{\hat{r}})^{a}(\partial_{\tau})^{b} \, - \, (\partial_{\hat{r}})^{b}(\partial_{\tau})^{a} \Big) \, ,
\end{align}
as can be checked easily using the metric in Eq.~\eqref{eq-ads-sand}. 
%Note that the Euclidean vector field $\xi_{E}^{a} \, = \, 2\pi \, (\partial_{\tau})^{a} \, $ is a Killing vector field of the Euclidean black hole in Eq.~\eqref{eq-ads-rind-2}. However, it is not a Killing vector field of the Euclidean bulk dual of the sandwiched state. Nevertheless, it is a killing vector field near $\tilde{B}_{(n)}$ as is evident from the metric in Eq.~\eqref{eq-ads-sand}. 
Furthermore, the boundary condition that the Euclidean bulk metric approaches the metric of $\M'$ near the asymptotic boundary guarantees that $\xi_{E}$ approaches the generator of the Euclidean time translation on the boundary. In other words, the vector field $\xi_{E}$ satisfies the Euclidean version of the boundary conditions in Eq.~\eqref{eq-eta-bc-intro} and Eq.~\eqref{eq-binormal}. %\tcb{(Relate this to the condition of the vector field whose conjugate Hamiltonian we are interested in. Do this after reviewing the results of OOguir et. al. in the introduction.)}
%\newline

Having established that the sandwiched state (for integer $n \ge 1$) has a bulk dual, we now use Eq.~\eqref{yy} to get
%This finishes our discussion of the Euclidean bulk dual of the vacuum and the sandwiched state. As discussed above, Eq.~\eqref{yy} implies 
\begin{align}
S_{\text{rel}}(\r ||\sigma) \, = \, H_{\xi}\left(\MM_{(n)}\right) \, - \, H_{\xi}\left(\MM_{\sigma}\right) \, , %
\end{align}
where $\xi \, = i\xi_{E} =  -2\pi \, \partial_{t}$ as in Eq.~\eqref{eq-eta-intro-2}. Moreover,  $\MM_{\sigma}$ and $\MM_{(n)} \, = \, \MM_{\r}$ denote the bulk spacetimes dual to the vacuum and the sandwiched state respectively. Now using Eq.~\eqref{eq-ren-rel-fin}, we conclude that the refined R\'enyi relative entropy between states $\rho$ and $\sigma$ is given by
\begin{align}
\S_{n} (\rho ||\sigma) \, = \, H_{\xi}\left(\MM_{(n)}\right) \, - \, H_{\xi}\left(\MM_{\sigma}\right) \, . \label{eq-sn-rel-EUC} 
\end{align}
%\newline

Our result in Eq.~\eqref{eq-sn-rel-EUC} needs to be contrasted with the holographic formula of the refined R\'enyi entropy which is defined in Eq.~\eqref{eq-Dong-Renyi}. According to the holographic formula derived in \cite{Dong-Renyi}, the refined R\'enyi entropy of a boundary state is equal to the area of a codimension-$2$ minimal area surface in some bulk spacetime other than the spacetime dual to the boundary state. In particular, the spacetime that one uses to compute refined R\'enyi entropy has to be constructed by introducing a cosmic brane and then accounting for its backreaction.  Recently, the boundary state dual to this backreacted geometry has been identified in \cite{Akers:2018fow,Dong:2018seb}. Our result in Eq.~\eqref{eq-sn-rel-EUC} is similar in the sense that the refined R\'enyi relative entropy between states $\rho$ and $\sigma$ is not related to the conserved charge in the bulk dual of the state $\rho$. In fact, it is related to the conserved charge in the bulk dual of a sandwiched state, $\r \, $. 
%\newline 

Despite this similarity, there are two crucial differences between our holographic formula for the refined R\'enyi relative entropy and the holographic formula for refined R\'enyi entropy derived in \cite{Dong-Renyi}. One is that the holographic formula for the refined R\'enyi entropy is valid for a general boundary theory and for arbitrary boundary subregions. On the other hand, our results apply only if the boundary theory is a CFT and the boundary subregion is spherical. Secondly, the bulk geometry that one needs to compute the refined R\'enyi entropy has a conical singularity at the location of the cosmic brane \cite{Dong-Renyi}, whereas the geometry that one needs for refined R\'enyi relative entropy, that is the bulk dual of the sandwiched state, is smooth. This is related to the fact that R\'enyi relative entropies, unlike R\'enyi entropies, are free of UV divergences\footnote{We thank Tom Hartman for pointing this out.}.
%One is that the geometry that one needs to compute the refined R\'enyi entropy has a conical singularity \cite{Dong-Renyi}, whereas the geometry that one needs for refined R\'enyi relative entropy, that is the bulk dual of the sandwiched state, is smooth. Secondly, the holographic formula for the refined R\'enyi entropy is valid for a general boundary theory and for arbitrary boundary subregions thereof. On the other hand, our results apply only if the boundary theory is a CFT and the boundary subregion is spherical.
%\newline

The holographic formula for refined R\'enyi relative entropy in Eq.~\eqref{eq-sn-rel-EUC} is one of the main results of this paper. In the following, we discuss the implications of this result and its relation with the our perturbative result in Sec.~(\ref{sec-pert-analysis}). 

\subsection{Positive energy theorems} \label{sec-pos-energy-thm}

The result of \cite{lashkari2016gravitational} that the holographic dual of the relative entropy is related to the conserved charge as in Eq.~\eqref{yy} led to interesting results. For example, the positive semi-definiteness of the relative entropy implies a positive energy theorem: the vacuum-subtracted conserved charge in the entanglement wedge of a spherical boundary region is non-negative. That is,
\begin{align}
H_{\xi}\left(\MM_{\rho}\right) \, - \, H_{\xi}\left(\MM_{\sigma}\right) \, \ge \, 0 \, . \label{eq-pos-energy-thm}
\end{align}
Note that Eq.~\eqref{eq-pos-energy-thm} is an infinite set of positivity conditions as it is true for all spherical boundary regions. We refer the interested readers to \cite{lashkari2016gravitational} for other gravitational constraints emerging from Eq.~\eqref{yy}.
%\newline

Now we discuss a different positive energy theorem that arises from the holographic dual of refined R\'enyi relative entropy in Eq.~\eqref{eq-sn-rel-EUC}. Recall from Eq.~\eqref{eq-RRRE-ineq} that the refined R\'enyi relative entropy (for $n \ge 1$) cannot be smaller than the relative entropy. That is,
\begin{align}
\S_{n} (\rho||\sigma) \, \ge \, S_{\text{rel}}\left(\rho || \sigma\right) \, \quad\quad \text{ for } n \, \ge \, 1 \, .
\end{align}
By combining this inequality with Eq.~\eqref{yy} and Eq.~\eqref{eq-sn-rel-EUC}, we get
\begin{align}
H_{\xi}\left(\MM_{(n)}\right) \, \ge \, H_{\xi}\left(\MM_{(1)}\right)  \, \quad\quad \text{ for } n \, \ge \, 1 \, , \label{eq-pos-energy-thm-2}
\end{align}
where $\MM_{(1)} \, = \, \MM_{\rho} \, $. Since the boundary conditions for the bulk dual of the sandwiched state are $\Z_{n}$-symmetric as shown in Fig.~(\ref{fig-euc-geom}), we conclude that making the boundary conditions symmetric in this manner cannot decrease the conserved charge in the entanglement wedge. Just like Eq.~\eqref{eq-pos-energy-thm}, Eq.~\eqref{eq-pos-energy-thm-2} is also an infinite set of conditions as it is valid for any spherical boundary regions.
%\newline

The positive energy condition in Eq.~\eqref{eq-pos-energy-thm-2} is interesting as it compares the conserved charges in two non-vacuum bulk geometries. In this sense, it is different from the usual positive energy theorems \cite{1982NuPhB,Woolgar:1994ar} which compare the energy in a ({non-vacuum}) spacetime with the energy in a vacuum spacetime. Hence, we consider  Eq.~\eqref{eq-pos-energy-thm-2} to be another nontrivial bulk consequence of a quantum information theoretic property.
%\newline

It is known that the sandwiched R\'enyi relative entropy monotonically increases with the R\'enyi parameter $n$, as stated in Eq.~\eqref{eq-mono-sand}. It is therefore natural to ask if the refined R\'enyi relative entropy obeys monotonicity as well. We do not have a definitive answer to this question, though we present a simple example in Appendix~(\ref{app-RRRE-thermal}) in which the refined R\'enyi relative entropy increases monotonically with the R\'enyi parameter. If it can be shown that the refined R\'enyi relative entropy monotonically increases with the R\'enyi parameter either in general or for the set of states $\rho$ and $\sigma$ in Eq.~\eqref{eq-rho-cft} and Eq.~\eqref{eq-vac} respectively, then this would imply a family of much stronger positive energy theorems than Eq.~\eqref{eq-pos-energy-thm-2}. We leave this question to future work. 

\subsection{Relation to the perturbative result}

As discussed earlier, the Euclidean bulk dual of the sandwiched state, $\r \, $, is the solution of the bulk Euclidean equations of motions for the metric and the scalar field subject to $\Z_{n}$-symmetric boundary conditions. Here, we denote the boundary sources by $\tilde{\lambda}_{(n)}$ and assume that the boundary sources are ``small,'' that is, we assume that the boundary sources can be written as $\tilde{\lambda}_{(n)} \, = \, \epsilon \, f_{(n)} \,$ where $\epsilon \ll 1 \,$. 
%\newline

The boundary conditions for the bulk scalar field demand that the scalar field, $\Phi$, can be written as
\begin{align}
\Phi \, = \, \epsilon \, \Phi^{(1)} \, + \, O(\epsilon^{2}) \, . \label{eq-bulk-sc-sand}
\end{align}
The bulk spacetime when $\epsilon \, = \, 0$ is the (Euclidean) AdS-Rindler spacetime. The correction to the metric has to be computed by solving the gravitational equations of motions sourced by the stress-energy tensor of the bulk scalar field. Since the stress-energy tensor is quadratic in the scalar field, it goes like $O(\epsilon^{2})$. Therefore, there is no backreaction at linear order in $\epsilon$ and hence, we can write the bulk metric as
\begin{align}
g \, = \, g^{(0)} \, + \, O(\epsilon^{2}) \, ,
\end{align}
where $g^{(0)}$ is the metric of the (Euclidean) AdS-Rindler spacetime. Moreover, this implies that $\Phi^{(1)}$ is the solution of the scalar field on a fixed AdS-Rindler spacetime with $\Z_{n}$-symmetric boundary conditions. 
%\newline

Now note that the conserved charge in the bulk dual of sandwiched state can be written as 
\begin{align}
H_{\xi}[ \, g \, ] \, = \, H_{\xi}[ \, g^{(0)} \, ] \, + \, \epsilon \, H_{\xi}^{\, (1)}[ \, g \, ] \,+ \, \epsilon^{2} \, H_{\xi}^{\, (2)}[ \, g \, ] \, + O(\epsilon^{3}) \, ,
\end{align}
Since $\xi$ is the killing vector field of the AdS-Rindler spacetime, we know from Eq.~\eqref{eq-pert-Heta-one} that $H_{\xi}^{\, (1)}$ vanishes identically. Moreover, from Eq.~\eqref{eq-pert-Heta-two}, we know that $H_{\xi}^{\, (2)}$ is given by
\begin{align}
H_{\xi}^{(2)} \, = \, \int_{\Sigma_{0}} \, \omega_{\phi}\left( \Phi^{(1)}, \lie_{\xi} \Phi^{(1)} \right) \, , \label{eq-pert-Heta-holo}
\end{align}
where we have used the fact that there is no correction to the metric at linear order in $\epsilon$. 
%\newline

Now combining Eq.~\eqref{eq-sn-rel-EUC} and Eq.~\eqref{eq-pert-Heta-holo}, we deduce that the refined R\'enyi relative entropy at the lowest order in the perturbation parameter is given by the symplectic flux of a scalar field, where the scalar field satisfies the $\Z_{n}$-symmetric boundary conditions. Hence, we recover Eq.~\eqref{eq-sn-cft-symp-fin} from Eq.~\eqref{eq-sn-rel-EUC}. However, this should not be considered a derivation of Eq.~\eqref{eq-sn-cft-symp-fin} as this analysis is only valid for CFT's with semi-classical holographic duals whereas Eq.~\eqref{eq-sn-cft-symp-fin} is valid for \textit{all} CFT's. 
%\newline

This finishes our discussion of the holographic dual of the refined R\'enyi relative entropy.

%In Sec.~(\ref{sec-pert-analysis}), we showed that the RRRE between a vacuum state and a perturbative state of a CFT reduced to a spherical region is equal to the symplectic flux of a scalar field through 

\section{Discussion} \label{sec-conc}

In this paper, we have introduced a quantity related to the sandwiched R\'enyi relative entropy, which we called the `refined' R\'enyi relative entropy, as defined in Eq.~\eqref{eq-rrre-def-intro}. Like the sandwiched R\'enyi relative entropy, this quantity is a one-parameter generalization of relative entropy. In particular, we found that the refined R\'enyi relative entropy between two states can be written as the relative entropy of two related states, as in Eq.~\eqref{eq-ren-rel-fin}. This identity relating the refined R\'enyi relative entropy with relative entropy played an important role in our analysis in this paper. 
%\newline

We derived a perturbative formula, Eq.~\eqref{eq-ren-rel-pert-fin}, for the refined R\'enyi relative entropy of two nearby states. Using this result, we studied the perturbative refined R\'enyi relative entropy when the reference state is the vacuum of a CFT reduced to a spherical region and the other state is a small perturbation thereof. We found that the perturbative refined R\'enyi relative entropy can be written as the symplectic flux of a scalar field through a Cauchy slice of the AdS-Rindler spacetime as in Eq.~\eqref{eq-sn-cft-symp-fin}. 
%\newline

We then studied the refined R\'enyi relative entropy for holographic CFT's. We argued that for a certain family of states reduced to a spherical region, the holographic dual of the refined R\'enyi relative entropy is related to certain conserved charges in the dual bulk spacetime. Combining this holographic result with an inequality in Eq.~\eqref{eq-RRRE-ineq} that the refined R\'enyi relative entropy must satisfy, we then proved certain `positive-energy' theorems in the asymptotically locally AdS spacetimes. 
%\newline

We now discuss some possible directions in which the present work can be extended.

\begin{enumerate}
	\item {\bf Data processing inequality} 
	
	As discussed in Sec.~(\ref{sec-rrre-def}), the relative entropy and the sandwiched R\'enyi relative entropy do not increase under a CPTP map \cite{Wilde-14,sand13,FL-13,beigi,MO-15,Lindblad1974}. This is known as the data-processing inequality, given in Eq.~\eqref{eq-dp-rel}. Since the partial trace operation is a CPTP map, the data-processing inequality implies that the relative and the sandwiched R\'enyi relative entropy for states of system $B$ is not greater than those for states of system $A$ if $B \subseteq A$, as stated in Eq.~\eqref{eq-mono-rel}. 
	%\newline
	
	A `good' measure of the distinguishability of two quantum states must satisfy the property that it does not increase under a partial trace operation. This is because the partial trace could in general result in the loss of some potentially distinguishing information, thus rendering distinguishability of the states more difficult. It would be interesting to investigate if the refined R\'enyi relative entropy satisfies a data-processing inequality. 
	
	\item {\bf Monotonicity in the R\'enyi parameter $n$}
	
	Recall from Eq.~\eqref{eq-mono-sand} that the sandwiched R\'enyi relative entropy monotonically increases with the R\'enyi parameter $n$ \cite{sand13,beigi}. We have derived a weaker version of this inequality for the refined R\'enyi relative entropy in Eqs.~\eqref{eq-RRRE-ineq}-\eqref{eq-RRRE-ineq-2}. We have used Eqs.~\eqref{eq-RRRE-ineq}-\eqref{eq-RRRE-ineq-2} in Sec.~(\ref{sec-pos-energy-thm}) to prove certain positive energy theorems in asymptotically locally AdS spacetimes. 
	%\newline
	
	It is not clear if the refined R\'enyi relative entropy monotonically increases with the R\'enyi parameter $n$ in general (although we showed it to be true in a simple example in Appendix~(\ref{app-RRRE-thermal}) .) If this were true, then the holographic formula for the refined R\'enyi relative entropy in Eq.~\eqref{eq-sn-rel-EUC} would lead to much stronger positive energy theorems than what we have found in this paper. Because of this, it would be worthwhile to explore if the refined R\'enyi relative entropy increases monotonically with $n$. 
	
	\item {\bf Generalization of quantum null energy condition (QNEC)}
	
	A QNEC is a local statement which puts a lower bound on the expectation value of the `null-null' component of the stress-energy tensor in an arbitrary state $\ket{\psi}$ of a quantum field theory in $R^{1,d-1}$ \cite{Bousso:2015mna}. More precisely, it states that $\langle T_{kk} \rangle$ for any null vector $k$ at any point $p$ satisfies
	\begin{align}
	\langle T_{kk} \rangle \, \ge \, \frac{1}{2\pi} \, S''(\rho) \, , \label{eq-qnec}
	\end{align}
	where $\rho$ is the state $\ket{\psi}$ reduced to any subregion $\Sigma_{p}$ such that the boundary of that subregion contains the point $p$. Moreover, $S(\rho)$ is the Von Neumann entropy of the state $\rho$ and prime denotes the infinitesimal variation of the subregion $\Sigma_{p}$ in the $k^{a}$ direction at point $p$. 
	%\newline
	
	In some specific cases, Eq.~\eqref{eq-qnec} can be written in terms of the relative entropy. In particular, if the boundary of the entangling region lies on a null hypersurface, Eq.~\eqref{eq-qnec} can be written as \cite{leichenauer2018energy}
	\begin{align}
	S''_{\text{rel}} (\rho||\sigma) \, \ge \, 0 \, , \label{eq-qnec-2}
	\end{align}
	where $\sigma$ is the reduced vacuum state. This follows from known results about the modular Hamiltonian when the entangling surface lies on a null plane \cite{Casini:2017roe}. 
	%\newline
	
	Recently, the sandwiched R\'enyi relative entropy between an excited state and a vacuum state of an arbitrary QFT reduced to a half-space was considered in \cite{Lashkari-sand}. It was found that the sandwiched R\'enyi relative entropy satisfies a bound analogous to Eq.~\eqref{eq-qnec-2}. It would be interesting to repeat this analysis for the refined R\'enyi relative entropy and see if it also satisfies such a bound. 
	
	% \item {\bf Shape deformations of refined R\'enyi relative entropy} using Eq.~\eqref{eq-ren-rel-pert-fin}
	
	\item {\bf Refined R\'enyi relative entropy under RG flow}
	
	In this project, we have only studied the refined R\'enyi relative entropy between the states of a conformal field theory. It would be interesting to study it away from critical points. Specifically, one may deform a CFT by a relevant operator and study the refined R\'enyi relative entropy between the vacuum state of the deformed theory and that of the original CFT. This calculation, to lowest order in the coupling, can be done using the general perturbative formula in Eq.~\eqref{eq-ren-rel-pert-fin}. It would be fascinating if one can gain some insights on RG flow, similar to those in \cite{add-ref-2}, by using inequalities in Eqs.~\eqref{eq-RRRE-ineq}-\eqref{eq-RRRE-ineq-2}.

	\item{\bf Further Inequalities for the refined R\'enyi relative entropy through Holography}
	
	It is also possible that further inequalities for the refined R\'enyi relative entropy can be discovered by exploiting holography. It would be interesting if the properties of the conserved charges in asymptotically locally AdS spacetimes can be used to find constraints obeyed by the refined R\'enyi relative entropy for holographic states that are not obeyed by that for all quantum states. These would be analogous to the constraints satisfied by the holographic entanglement entropy known as the holographic entropy cone \cite{hayden2013holographic,bao2015holographic}.
	%Because it is dual to a geometric object, it is possible that this geometric nature can be exploited to find constraints obeyed by the refined R\'enyi relative entropy for holographic states that are not obeyed by that for all quantum states, in a program analogous to that of the holographic entanglement entropy program, in particular that of the holographic entropy cone \cite{bao2015holographic, hayden2013holographic}.
	
	\item{\bf Generalization of JLMS formula}
	
	It is known that the relative entropy of two nearby boundary states is equal to the relative entropy of the corresponding bulk states \cite{JLMS}. This holographic relation was an important ingredient in proofs of bulk reconstruction \cite{dong2016reconstruction, faulkner2017bulk, cotler2017entanglement}. It would be interesting to explore if an analogous relation holds for refined R\'enyi relative entropy and what its implications are for bulk reconstruction.
	
	% Finally, it is possible that a version of the JLMS formula \cite{JLMS} can also hold for the refined R\'enyi relative entropy. It would be interesting to explore the duality between bulk and boundary RRRE's in the same way as for bulk and boundary relative entropies, and to study their potential role in studying bulk reconstruction.
	
\end{enumerate}

\section*{Acknowledgements}
We would like to thank Nima Afkhami-Jeddi, Ahmed Almheiri, Tom Faulkner, Tom Hartman, Nima Lashkari, and Pratik Rath for useful discussions during the course of this work. NB is supported by the National Science Foundation under grant number 82248-13067-44-PHPXH, by the Department of Energy under grant number DE-SC0019380, and by New York State Urban Development Corporation – Empire State Development – contract no. AA289. MM is supported by the US Department of Energy under grant number DE-SC0014123. 

\appendix

%\section{The relation between relative and refined R\'enyi relative entropies} \label{app-sn}
\section{Refined R\'enyi relative entropy as relative entropy} \label{app-sn}

In Sec.~(\ref{sec-rrre-def}), we claimed that the refined R\'enyi relative entropy between two states can be written as the relative entropy between two related states as in Eq.~\eqref{eq-ren-rel-fin}. We found this identity very useful in deriving a general formula for the perturbative refined R\'enyi relative entropy, given in Eq.~\eqref{eq-ren-rel-pert-fin}. We applied this general perturbative formula in Sec.~(\ref{sec-pert-analysis}) to show that the refined R\'enyi relative entropy between two states of \textit{any} CFT reduced to a spherical region can be written as the symplectic flux of a scalar field through a Cauchy slice of the AdS-Rindler wedge as in Eq.~\eqref{eq-sn-cft-symp-fin}. Moreover, we combined the identity in Eq.~\eqref{eq-ren-rel-fin} with Eq.~\eqref{yy} in Sec.~(\ref{sec-non-pert-analysis}) to deduce the holographic dual of the refined R\'enyi relative entropy. 
%\newline

In this appendix, we present a derivation of the Eq.~\eqref{eq-ren-rel-fin}. We start with the definition of refined R\'enyi relative entropy
\begin{align}
\S_{n} (\rho||\sigma) \, \equiv \, n^{2} \, \partial_{n} \left( \frac{n-1}{n} \, S_{n} (\rho||\sigma) \right) \, , \label{eq-mod-ren-app}
\end{align}
where 
\begin{align}
S_{n} (\rho || \sigma) \, \equiv \, \frac{1}{n-1} \, \log\tr \left\{ \Big(\sigma^{\frac{1-n}{2n}} \, \rho \, \sigma^{\frac{1-n}{2n}} \Big)^{n} \right\} \, , \label{eq-ren-rel-app}
\end{align}
is the sandwiched R\'enyi relative entropy. Combining the above two expressions, we get
\begin{align}
\S_{n} (\rho||\sigma) \, \equiv \, n^{2} \, \partial_{n} \left( \frac{1}{n} \, \log \, \tr \, \hat{\rho}_{(n)}^{\, n} \right) \, , \label{eq-sn-app-1}
\end{align}
where we have defined a Hermitian matrix 
\begin{align}
\hat{\rho}_{(n)} \, \equiv \, \sigma^{\frac{1-n}{n}} \, \rho \, \sigma^{\frac{1-n}{n}} \, . \label{eq-sand-int-app}
\end{align}
Taking the derivative w.r.t $n$ in Eq.~\eqref{eq-sn-app-1}, we get
\begin{align}
\S_{n} (\rho||\sigma) \, = \, - \, \log \, \tr \, \hat{\rho}_{(n)}^{\, n} \, + \,   \frac{n}{\tr \, \hat{\rho}_{(n)}^{\, n}} \, \partial_{n} \, \tr \, \hat{\rho}_{(n)}^{\, n}  \, . \label{eq-sn-app-4}
\end{align}
Now using
\begin{align}
\partial_{n} \, \tr \, \hat{\rho}_{(n)}^{\, n} \, = \, \tr \left( \hat{\rho}_{(n)}^{\, n} \, \cdot \, \log \hat{\rho}_{(n)} \right) \, + \, n \, \tr \left( \hat{\rho}_{(n)}^{\, n-1} \, \cdot \, \partial_{n} \hat{\rho}_{(n)} \right) \, 
\end{align}
and
\begin{align}
\partial_{n} \hat{\rho}_{(n)} \, = \, - \frac{1}{2 n^{2}} \, \left( \log\sigma \cdot \hat{\rho}_{(n)} \, + \,  \hat{\rho}_{(n)} \cdot \log\sigma \right) \, ,
\end{align}
we write Eq.~\eqref{eq-sn-app-4} as
\begin{align}
\S_{n} (\rho||\sigma) \, = \, - \, \log \, \tr \, \hat{\rho}_{(n)}^{\, n} \, + \, \frac{n}{\tr \, \hat{\rho}_{(n)}^{\, n}} \, \tr \left( \hat{\rho}_{(n)}^{\, n} \, \cdot \, \log \hat{\rho}_{(n)} \right) \, - \, \frac{1}{\tr \, \hat{\rho}_{(n)}^{\, n}} \, \tr \left( \hat{\rho}_{(n)}^{\, n} \, \cdot \, \log \sigma \right)  \, . \label{eq-sn-app-2}
\end{align}
We simplify this result and write it as
\begin{align}
\S_{n} (\rho||\sigma) \, = \, \tr \left( \rho_{(n)} \, \log \rho_{(n)} \right) \, - \, \tr \left( \rho_{(n)} \, \log \sigma \right)  \, , \label{eq-sn-app-3}
\end{align}
where 
\begin{align}
\r \, \equiv \, \frac{\hat{\rho}_{(n)}^{\, n}}{\tr \hat{\rho}_{(n)}^{\, n} } \, 
\end{align}
is precisely the sandwiched state defined in Eq.~\eqref{eq-sand-st}. Now the expression in the r.h.s of Eq.~\eqref{eq-sn-app-3} is the same as the definition of the relative entropy given in Eq.~\eqref{eq-re-def}. However, to relate the r.h.s of Eq.~\eqref{eq-sn-app-3} with the relative entropy, we first need to show that the sandwiched state $\r$ is a valid denisty matrix. In other words, it is a positive semi-definite matrix with unit trace. Note that, by construction, $\r$ has unit trace. Moreover, the fact that $\r$ is positive semi-definite follows from the fact that the matrix $\hat{\rho}_{(n)}$ in Eq.~\eqref{eq-sand-int-app} is positive semi-definite. To see this, note that the expectation value of $\hat{\rho}_{(n)}$  in any arbitrary state $\ket{\psi}$ is non-negative:
\begin{align}
\bra{\psi}\hat{\rho}_{(n)} \ket{\psi} \, = \, \bra{\psi'} \rho \ket{\psi'} \, \ge \, 0 \, ,
\end{align}
where $\ket{\psi'} \, \equiv \, \sigma^{\frac{1-n}{n}} \ket{\psi} \, $. Since, the sandwiched state is positive semi-definite matrix with unit trace, it is a valid density matrix. Hence, we can write the r.h.s of Eq.~\eqref{eq-sn-app-3} as relative entropy. More precisely,
\begin{align}
\S_{n} (\rho||\sigma) \, = \,\, S_{\text{rel}} \left(\r||\sigma\right) \, .
\end{align}
This finishes our derivation of Eq.~\eqref{eq-ren-rel-fin}.

\section{Conformal transformation and state $\rho$} \label{app-rho}

In this paper, we considered states that are obtained by acting on the vacuum of a CFT with a single smeared operator. In Sec.~(\ref{sec-setup}), we stated that these states reduced to a spherical region are related to some states on the hyperbolic space by a unitary transformation. We claimed that the precise form of these reduced states is given in eq.~\eqref{eq-rho-cft}. In this appendix, we present a derivation of Eq.~\eqref{eq-rho-cft}. %\tcb{(Add details after finishing the non-perturbative section.)}
%\newline

Consider a CFT with Euclidean action $I_{0}$ on $R^{d}$. We work in the coordinate systems in which the metric of $R^{d}$ is
\begin{align}
ds^{2} \, = \, d t_{E}^{2} + dr^{2} + r^{2} \, d \chi^{2}_{d-2} \, , \label{eq-met-Rd-app}
\end{align}
where $d \chi^{2}_{d-2}$ is the metric of a $(d-2)$-dimensional sphere of unit radius. Now consider a state $\ket{\Psi(-t_{E 0})} \, \equiv \, \Psi(-t_{E 0}) \ket{\Omega}$, where $\ket{\Omega}$ is the vacuum state of the CFT and $\Psi(-t_{E 0})$ is a  local operator $\Psi(x)$ smeared in a small neighborhood around $t_{E} \, = - t_{E 0} \, < \, 0 \, $. Note that due to translation symmetry, this state is independent of where we insert the operator $\Psi$ at $t_{E} = -t_{E 0}$ slice. This means we can take the operator to be smeared in a small region around $(t_{E} , r) = (-t_{E 0} , 0) \,$. For simplicity, we assume that the operator $\Psi$ is Hermitian. 
%\newline

Now let $\rho$ be the state $\ket{\Psi(-t_{E0})}$ reduced to a spherical region $B$ of radius $R$. We take the region $B$ to be given by 
\begin{align}
B: \, \quad t_{E} = 0 \, \quad \text{and } \, \quad r  \le R  \, . \label{eq-reg-B-app}
\end{align}
The density matrix of $\rho$ can can be written as a Euclidean path integral over $R^{d}$ with open cuts just above and below the region $B$ and with the insertion of $\Psi(-t_{E0})$ and $\Psi(t_{E0}) \, $. That is, the matrix elements of $\rho$ (up to a normalization constant) are given by
\begin{align}
\bra{\phi_{-}}\rho\ket{\phi_{+}} \, \sim \, \int_{\Phi(B_{\pm}) = \phi_{\pm}} \, D\Phi \, e^{-I_{0}[\Phi]} \,\,\, \Psi(-t_{E0}) \, \Psi(t_{E0}) \, , \label{eq-den-app}
\end{align}
where we have imposed the boundary conditions at the open cuts 
\begin{align}
B_{\pm}: \, \quad t_{E} = 0^{\pm} \, \quad \text{and } \, \quad r  \le R  \, . 
\end{align}
%\newline

Now note that $R^{d}$ can be conformally mapped to $\M' \, \equiv \, S^{1} \times \hp$. This conformal transformation is given by \cite{Rosenhaus:2014woa}
\begin{align}
t_{E} \, =& \, R \, \frac{\sin (\tau/R)}{\cosh u + \cos (\tau/R) } \, \quad\quad\quad\quad  r \, = \, R \, \frac{\sinh u}{\cosh u + \cos (\tau/R) } \, , \label{eq-ct-r-app}
\end{align}
where $0 \le \tau \le 2\pi R$ and $0 \le u < \infty$. Under this coordinate transformation, the metric of $R^{d}$ in Eq.~\eqref{eq-met-Rd-app} becomes
\begin{align}
ds^{2} \, = \, \big( \cosh u + \cos (\tau/R) \big)^{-2} \, \Big( d\tau^{2} \, + R^{2} \left( du^{2} + \sinh^{2}u \, d\chi^{2}_{d-2} \right) \Big) \,  ,
\end{align}
which is the same as the metric of $\M' \, = \, S^{1} \times \hp$, up to a conformal factor.
%\begin{align}
%\chi(\theta,u) \, = \, \left( \cosh u + \cos (\theta/R) %\right)^{-1} \, .
%\end{align}
%\newline

Using the conformal transformation in Eq.~\eqref{eq-ct-r-app}, the matrix element in Eq.~\eqref{eq-den-app} can be written as a path integral over $\M'$. Note that the branch cut $B_{+}$ maps to $\tau = 0$ and $0 \le u < \infty$ whereas the branch cut $B_{-}$ maps to $\tau = 2\pi R$ and $0 \le u < \infty$. Moreover, the points $(t_{E} = \pm t_{E0}$, $r=0$) map to $(\tau = \pi R \mp \tau_{0} , u =0)\, $, where %\tcr{(What is $\Omega$ here?)} where \tcb{(Please mention after ($74$) that it is the metric on $d-2$-dimensional sphere of unit radius. Also, we need to replace this $\Omega$ with some other symbol as $\Omega$ also denotes the vacuum.)}
\begin{align}
\tau_{0} \, = \, 2 R \arctan(R/t_{E0}) \, .
\end{align}
Hence, the matrix element of $\rho$ can be written as 
\begin{align}
\bra{\phi_{-}}\rho\ket{\phi_{+}} \, \sim& \,  \int^{\Phi(\tau=2\pi R)=\tilde{\phi}_{-}}_{\Phi(\tau=0)=\tilde{\phi}_{+}} \, D\Phi \, e^{-I_{\text{0}}[\Phi]} \,\,\, \tilde{\Psi}(\tau = \pi R + \tau_{0}) \, \tilde{\Psi}(\tau = \pi R - \tau_{0}) \, , \label{eq-den-two-app} 
\end{align}
where $|\tilde{\phi}\rangle \equiv U\ket{\phi}$ and $\tilde{\Psi} \equiv U \Psi U^{\dagger}$ are the unitary transformations of the states and operators under the aforementioned conformal transformation. This finishes our derivation of Eq.~\eqref{eq-rho-cft-pi}. 
%\newline

Now note that we can write the path integral in Eq.~\eqref{eq-den-two-app} in operator language. In the Schrodinger picture, we get
\begin{align}
\bra{\phi_{-}}\rho\ket{\phi_{+}} \, \sim \, \langle \tilde{\phi}_{-}| \, e^{-(\pi R - \tau_{0}) H} \,\, \tilde{\Psi}(0) \,\, e^{-2\tau_{0} H} \, \tilde{\Psi}(0) \,\, e^{-(\pi R - \tau_{0}) H} \, |\tilde{\phi}_{+}\rangle \, ,
\end{align}
whereas in the Heisenberg picture, we get
\begin{align}
\bra{\phi_{-}}\rho\ket{\phi_{+}}\, \sim \, \langle \tilde{\phi}_{-}| \, e^{-2\pi R H} \,\, \tilde{\Psi}(\pi R + \tau_{0}) \,\, \tilde{\Psi}(\pi R - \tau_{0}) \, |\tilde{\phi}_{+}\rangle \, ,
\end{align}
where $H$ is the Hamiltonian of the CFT on the hyperbolic space. Equivalently, we can write the above expression as
\begin{align}
U \, \rho \, U^{\dagger} \, \sim \,  e^{-2\pi R H} \,\, \tilde{\Psi}(\pi R + \tau_{0}) \,\, \tilde{\Psi}(\pi R - \tau_{0})  \, . \label{eq-rho-app-int}
\end{align}
%\newline

In a simple case when the operator $\Psi$ is an identity operator, the state $\rho$ reduces to the vacuum state which we denote by $\sigma$. Therefore, from Eq.~\eqref{eq-rho-app-int}, we deduce that the state $\sigma$ is 
\begin{align}
U \, \sigma \, U^{\dagger} \, \sim \,  e^{-2\pi R H} \, \equiv \, \tilde{\sigma} \, , \label{eq-vac-app-non}
\end{align}
which is the thermal state on the hyperbolic space, $\tilde{\sigma} \, $, up to a unitary transformation. Note that Eq.~\eqref{eq-vac-app-non} is the same as Eq.~\eqref{eq-vac} up to a normalization constant. Moreover, inserting Eq.~\eqref{eq-vac-app-non} in Eq.~\eqref{eq-rho-app-int} yields
\begin{align}
U \, \rho \, U^{\dagger} \, \sim \,  \tilde{\sigma} \, \tilde{\Psi}(\pi R + \tau_{0}) \,\, \tilde{\Psi}(\pi R - \tau_{0})  \, ,
\end{align}
which is the same as Eq.~\eqref{eq-rho-cft} up to a normalization constant. This finishes the derivation of Eq.~\eqref{eq-rho-cft} and our discussion of the states that we considered in this paper.

\section{Refined R\'enyi relative entropy for thermal states} \label{app-RRRE-thermal}

In this appendix, we study the refined R\'enyi relative entropy between two thermal states at different temperatures and show that it monotonically increases with the R\'enyi parameter $n$. Consider a quantum system with Hamiltonian, $H$, and two thermal states $\rho_{(0)}$ and $\rho_{(1)}$ such that 
\begin{align}
\rho_{(i)} \, = \, \frac{ \, e^{-\beta_{i} \, H \,} }{ Z(\beta_{i}) } \, \quad\quad\quad \text{for } \quad\quad i \, = \, \{1,2\} \, ,
\end{align}
where 
\begin{align}
Z(\beta) \, \equiv \, \tr e^{-\beta \, H} \,  \label{eq-part-fn-app}
\end{align}
is the thermal partition function. 
%\newline

Now using Eq.~\eqref{eq-ren-rel-fin}, the refined R\'enyi relative entropy between $\rho_{1}$ and $\rho_{0}$ is given by
\begin{align}
\S_{n} (\rho_{(1)}||\rho_{(0)}) \, = \,\, S_{\text{rel}} \left(\r||\rho_{(0)}\right) \, , \label{eq-th-rrre-app}
\end{align}
where the sandwiched state, $\r$, is 
\begin{align}
\r \, = \, \frac{ \, \Big({\rho_{(0)}^{\frac{1-n}{2n}} \, \rho_{1} \, \rho_{(0)}^{\frac{1-n}{2n}}}\Big)^{n} \, }{ \, \tr \Big({\rho_{(0)}^{\frac{1-n}{2n}} \, \rho_{1} \, \rho_{(0)}^{\frac{1-n}{2n}}}\Big)^{n} \, } \, .
\end{align}
Since, $\rho_{(0)}$ and $\rho_{(1)}$ commute, we can simplify $\r$ and write it as
\begin{align}
\r \, = \, \frac{ \, e^{-\beta_{n} \, H \,} }{ Z(\beta_{n}) } \, ,
\end{align}
where 
\begin{align}
\beta_{n} \, = \, \beta_{0} \, + \, n \, (\beta_{1} - \beta_{0} ) \, . \label{eq-beta-n-app}
\end{align}
We consider the case when $\beta_{1} \, > \, \frac{n-1}{n}\beta_{0}$. This ensures that $\beta_{n} > 0$. Hence, we can think of $\r$ as a thermal state with $n$ dependent temperature. 
%\newline

Now using the definition of the relative entropy in Eq.~\eqref{eq-re-def}, we write Eq.~\eqref{eq-th-rrre-app} as
\begin{align}
\S_{n} (\rho_{(1)}||\rho_{(0)}) \, = \, \tr \left( \r \, \log \r \right) \, + \, \beta_{0} \, \tr \left( \r \, H \right) \, + \, \log Z(\beta_{0}) \, , 
\end{align}
or equivalently as
\begin{align}
\S_{n} (\rho_{(1)}||\rho_{(0)}) \, = \, \beta_{0} \, E(\beta_{n}) \, - \, S(\beta_{n}) \, + \, \log Z(\beta_{0}) \, ,  \label{eq-th-rrre-app-2}
\end{align}
where $E(\beta)$ and $S(\beta)$ are the average thermal energy and the thermal entropy of the system at temperature $1/\beta$. 
%\newline

Next we take the derivative of Eq.~\eqref{eq-th-rrre-app-2} w.r.t $n$ and get
\begin{align}
\partial_{n} \, \S_{n} (\rho_{(1)}||\rho_{(0)}) \, =& \, \beta_{0} \, \partial_{n} \, E(\beta_{n}) \, - \, \partial_{n} \, S(\beta_{n}) \, . %\\
%=& \, (\beta_{0} \, - \, \beta_{n})  \, \partial_{n} \, E(\beta_{n}) \, ,
\end{align}
Now using the first law of thermodynamics, $\delta E(\beta) \, = \, \beta^{-1} \, \delta S(\beta) \,$, and using Eq.~\eqref{eq-beta-n-app}, we simplify this to get
\begin{align}
\partial_{n} \, \S_{n} (\rho_{(1)}||\rho_{(0)}) \, = \, - \, n \, (\beta_{1} - \beta_{0}) \, \partial_{n} \, E(\beta_{n})  \, = \, - \, n \, (\beta_{1} \, - \, \beta_{0})^{2}  \,\,  \Big(\partial_{\beta} \, E(\beta)\Big)\Big|_{\beta = \beta_{n}} \, . \label{eq-th-RRRE-C}
\end{align}
Now note that $\partial_{\beta} E(\beta) \, \le \, 0 \,$. This can be seen as follows:
\begin{align}
\partial_{\beta} \, E(\beta) \, = \, \partial_{\beta} \, \langle H \rangle_{\beta} \, = \, \partial_{\beta} \, \frac{ \,  \tr (e^{-\beta \, H} \, H) \, }{ \, \tr (e^{-\beta \, H}) } \, = \, - \, \langle H^{2} \rangle_{\beta} + \langle H \rangle_{\beta}^{2} \, \le \, 0 \, . 
\end{align}
Combining this observation with Eq.~\eqref{eq-th-RRRE-C}, we deduce that
\begin{align}
\partial_{n} \, \S_{n} (\rho_{(1)}||\rho_{(0)}) \, \ge \, 0 \, .
\end{align}
Hence, the refined R\'enyi relative entropy between two thermal states of the same system monotonically increases with the R\'enyi parameter.

\bibliographystyle{utcaps}
\bibliography{wormholes-arXiv_v1}

\providecommand{\href}[2]{#2}\begingroup\raggedright\begin{thebibliography}{10}

\bibitem{Faulk-GR-entanglement}
T.~Faulkner, F.~M. Haehl, E.~Hijano, O.~Parrikar, C.~Rabideau, and
  M.~Van~Raamsdonk, ``{Nonlinear Gravity from Entanglement in Conformal Field
  Theories},'' \href{http://dx.doi.org/10.1007/JHEP08(2017)057}{{\em JHEP} {\bf
  08} (2017)  057},
\href{http://arxiv.org/abs/1705.03026}{{\tt arXiv:1705.03026 [hep-th]}}.
%%CITATION = ARXIV:1705.03026;%%.

\bibitem{lashkari2016gravitational}
N.~Lashkari, J.~Lin, H.~Ooguri, B.~Stoica, and M.~Van~Raamsdonk,
  ``Gravitational positive energy theorems from information inequalities,''
  {\em Progress of Theoretical and Experimental Physics} {\bf 2016} (2016)
  no.~12, .

\bibitem{Casini:2008cr}
H.~Casini, ``{Relative entropy and the Bekenstein bound},''
  \href{http://dx.doi.org/10.1088/0264-9381/25/20/205021}{{\em Class. Quant.
  Grav.} {\bf 25} (2008)  205021},
\href{http://arxiv.org/abs/0804.2182}{{\tt arXiv:0804.2182 [hep-th]}}.
%%CITATION = ARXIV:0804.2182;%%.

\bibitem{Longo:2018zib}
R.~Longo and F.~Xu, ``{Comment on the Bekenstein bound},''
  \href{http://dx.doi.org/10.1016/j.geomphys.2018.03.004}{{\em J. Geom. Phys.}
  {\bf 130} (2018)  113--120},
\href{http://arxiv.org/abs/1802.07184}{{\tt arXiv:1802.07184 [math-ph]}}.
%%CITATION = ARXIV:1802.07184;%%.

\bibitem{Wall:2010cj}
A.~C. Wall, ``{A Proof of the generalized second law for rapidly-evolving
  Rindler horizons},'' \href{http://dx.doi.org/10.1103/PhysRevD.82.124019}{{\em
  Phys. Rev.} {\bf D82} (2010)  124019},
\href{http://arxiv.org/abs/1007.1493}{{\tt arXiv:1007.1493 [gr-qc]}}.
%%CITATION = ARXIV:1007.1493;%%.

\bibitem{wall2012proof}
A.~C. Wall, ``Proof of the generalized second law for rapidly changing fields
  and arbitrary horizon slices,'' {\em Physical Review D} {\bf 85} (2012)
  no.~10, 104049.

\bibitem{Bousso:2014sda}
R.~Bousso, H.~Casini, Z.~Fisher, and J.~Maldacena, ``{Proof of a Quantum Bousso
  Bound},'' \href{http://dx.doi.org/10.1103/PhysRevD.90.044002}{{\em Phys.
  Rev.} {\bf D90} (2014) no.~4, 044002},
\href{http://arxiv.org/abs/1404.5635}{{\tt arXiv:1404.5635 [hep-th]}}.
%%CITATION = ARXIV:1404.5635;%%.

\bibitem{Bousso:2014uxa}
R.~Bousso, H.~Casini, Z.~Fisher, and J.~Maldacena, ``{Entropy on a null surface
  for interacting quantum field theories and the Bousso bound},''
  \href{http://dx.doi.org/10.1103/PhysRevD.91.084030}{{\em Phys. Rev.} {\bf
  D91} (2015) no.~8, 084030},
\href{http://arxiv.org/abs/1406.4545}{{\tt arXiv:1406.4545 [hep-th]}}.
%%CITATION = ARXIV:1406.4545;%%.

\bibitem{faulkner2016modular}
T.~Faulkner, R.~G. Leigh, O.~Parrikar, and H.~Wang, ``Modular Hamiltonians for
  deformed half-spaces and the averaged null energy condition,'' {\em Journal
  of High Energy Physics} {\bf 2016} (2016) no.~9, 38.

\bibitem{koeller2018local}
J.~Koeller, S.~Leichenauer, A.~Levine, and A.~Shahbazi-Moghaddam, ``Local
  modular Hamiltonians from the quantum null energy condition,'' {\em Physical
  Review D} {\bf 97} (2018) no.~6, 065011.

\bibitem{leichenauer2018energy}
S.~Leichenauer, A.~Levine, and A.~Shahbazi-Moghaddam, ``Energy is
  entanglement,'' {\em arXiv preprint arXiv:1802.02584} (2018)  .

\bibitem{Ceyhan:2018zfg}
F.~Ceyhan and T.~Faulkner, ``{Recovering the QNEC from the ANEC},''
\href{http://arxiv.org/abs/1812.04683}{{\tt arXiv:1812.04683 [hep-th]}}.
%%CITATION = ARXIV:1812.04683;%%.

\bibitem{add-ref-3}
H.~Casini, E.~Teste, and G.~Torroba, ``{Relative entropy and the RG flow},''
  \href{http://dx.doi.org/10.1007/JHEP03(2017)089}{{\em JHEP} {\bf 03} (2017)
  089},
\href{http://arxiv.org/abs/1611.00016}{{\tt arXiv:1611.00016 [hep-th]}}.
%%CITATION = ARXIV:1611.00016;%%.

\bibitem{BO}
N.~Bao and H.~Ooguri, ``{Distinguishability of black hole microstates},''
  \href{http://dx.doi.org/10.1103/PhysRevD.96.066017}{{\em Phys. Rev.} {\bf
  D96} (2017) no.~6, 066017},
\href{http://arxiv.org/abs/1705.07943}{{\tt arXiv:1705.07943 [hep-th]}}.
%%CITATION = ARXIV:1705.07943;%%.

\bibitem{JLMS}
D.~L. Jafferis, A.~Lewkowycz, J.~Maldacena, and S.~J. Suh, ``{Relative entropy
  equals bulk relative entropy},''
  \href{http://dx.doi.org/10.1007/JHEP06(2016)004}{{\em JHEP} {\bf 06} (2016)
  004},
\href{http://arxiv.org/abs/1512.06431}{{\tt arXiv:1512.06431 [hep-th]}}.
%%CITATION = ARXIV:1512.06431;%%.

\bibitem{dong2016reconstruction}
X.~Dong, D.~Harlow, and A.~C. Wall, ``Reconstruction of bulk operators within
  the entanglement wedge in gauge-gravity duality,'' {\em Physical review
  letters} {\bf 117} (2016) no.~2, 021601.

\bibitem{faulkner2017bulk}
T.~Faulkner and A.~Lewkowycz, ``Bulk locality from modular flow,'' {\em Journal
  of High Energy Physics} {\bf 2017} (2017) no.~7, 151.

\bibitem{cotler2017entanglement}
J.~Cotler, P.~Hayden, G.~Salton, B.~Swingle, and M.~Walter, ``Entanglement
  wedge reconstruction via universal recovery channels,'' {\em arXiv preprint
  arXiv:1704.05839} (2017)  .

\bibitem{Lashkari:2013koa}
N.~Lashkari, M.~B. McDermott, and M.~Van~Raamsdonk, ``{Gravitational dynamics
  from entanglement 'thermodynamics'},''
  \href{http://dx.doi.org/10.1007/JHEP04(2014)195}{{\em JHEP} {\bf 04} (2014)
  195},
\href{http://arxiv.org/abs/1308.3716}{{\tt arXiv:1308.3716 [hep-th]}}.
%%CITATION = ARXIV:1308.3716;%%.

\bibitem{Faulkner:2013ica}
T.~Faulkner, M.~Guica, T.~Hartman, R.~C. Myers, and M.~Van~Raamsdonk,
  ``{Gravitation from Entanglement in Holographic CFTs},''
  \href{http://dx.doi.org/10.1007/JHEP03(2014)051}{{\em JHEP} {\bf 03} (2014)
  051},
\href{http://arxiv.org/abs/1312.7856}{{\tt arXiv:1312.7856 [hep-th]}}.
%%CITATION = ARXIV:1312.7856;%%.

\bibitem{Banerjee:2014oaa}
S.~Banerjee, A.~Bhattacharyya, A.~Kaviraj, K.~Sen, and A.~Sinha,
  ``{Constraining gravity using entanglement in AdS/CFT},''
  \href{http://dx.doi.org/10.1007/JHEP05(2014)029}{{\em JHEP} {\bf 05} (2014)
  029},
\href{http://arxiv.org/abs/1401.5089}{{\tt arXiv:1401.5089 [hep-th]}}.
%%CITATION = ARXIV:1401.5089;%%.

\bibitem{Swingle:2014uza}
B.~Swingle and M.~Van~Raamsdonk, ``{Universality of Gravity from
  Entanglement},''
\href{http://arxiv.org/abs/1405.2933}{{\tt arXiv:1405.2933 [hep-th]}}.
%%CITATION = ARXIV:1405.2933;%%.

\bibitem{Banerjee:2014ozp}
S.~Banerjee, A.~Kaviraj, and A.~Sinha, ``{Nonlinear constraints on gravity from
  entanglement},'' \href{http://dx.doi.org/10.1088/0264-9381/32/6/065006}{{\em
  Class. Quant. Grav.} {\bf 32} (2015) no.~6, 065006},
\href{http://arxiv.org/abs/1405.3743}{{\tt arXiv:1405.3743 [hep-th]}}.
%%CITATION = ARXIV:1405.3743;%%.

\bibitem{lin2014tomography}
J.~Lin, M.~Marcolli, H.~Ooguri, and B.~Stoica, ``Tomography from
  entanglement,'' {\em arXiv preprint arXiv:1412.1879} (2014)  .

\bibitem{lashkari2015inviolable}
N.~Lashkari, C.~Rabideau, P.~Sabella-Garnier, and M.~Van~Raamsdonk,
  ``Inviolable energy conditions from entanglement inequalities,'' {\em Journal
  of High Energy Physics} {\bf 2015} (2015) no.~6, 67.

\bibitem{Lashkari:2015hha}
N.~Lashkari and M.~Van~Raamsdonk, ``{Canonical Energy is Quantum Fisher
  Information},'' \href{http://dx.doi.org/10.1007/JHEP04(2016)153}{{\em JHEP}
  {\bf 04} (2016)  153},
\href{http://arxiv.org/abs/1508.00897}{{\tt arXiv:1508.00897 [hep-th]}}.
%%CITATION = ARXIV:1508.00897;%%.

\bibitem{Faulkner:2014jva}
T.~Faulkner, ``{Bulk Emergence and the RG Flow of Entanglement Entropy},''
  \href{http://dx.doi.org/10.1007/JHEP05(2015)033}{{\em JHEP} {\bf 05} (2015)
  033},
\href{http://arxiv.org/abs/1412.5648}{{\tt arXiv:1412.5648 [hep-th]}}.
%%CITATION = ARXIV:1412.5648;%%.

\bibitem{Lashkari:2014yva}
N.~Lashkari, ``{Relative Entropies in Conformal Field Theory},''
  \href{http://dx.doi.org/10.1103/PhysRevLett.113.051602}{{\em Phys. Rev.
  Lett.} {\bf 113} (2014)  051602},
\href{http://arxiv.org/abs/1404.3216}{{\tt arXiv:1404.3216 [hep-th]}}.
%%CITATION = ARXIV:1404.3216;%%.

\bibitem{Blanco:2013joa}
D.~D. Blanco, H.~Casini, L.-Y. Hung, and R.~C. Myers, ``{Relative Entropy and
  Holography},'' \href{http://dx.doi.org/10.1007/JHEP08(2013)060}{{\em JHEP}
  {\bf 08} (2013)  060},
\href{http://arxiv.org/abs/1305.3182}{{\tt arXiv:1305.3182 [hep-th]}}.
%%CITATION = ARXIV:1305.3182;%%.

\bibitem{Rosenhaus:2014woa}
V.~Rosenhaus and M.~Smolkin, ``{Entanglement Entropy: A Perturbative
  Calculation},'' \href{http://dx.doi.org/10.1007/JHEP12(2014)179}{{\em JHEP}
  {\bf 12} (2014)  179},
\href{http://arxiv.org/abs/1403.3733}{{\tt arXiv:1403.3733 [hep-th]}}.
%%CITATION = ARXIV:1403.3733;%%.

\bibitem{Rosenhaus:2014ula}
V.~Rosenhaus and M.~Smolkin, ``{Entanglement entropy, planar surfaces, and
  spectral functions},'' \href{http://dx.doi.org/10.1007/JHEP09(2014)119}{{\em
  JHEP} {\bf 09} (2014)  119},
\href{http://arxiv.org/abs/1407.2891}{{\tt arXiv:1407.2891 [hep-th]}}.
%%CITATION = ARXIV:1407.2891;%%.

\bibitem{Allais:2014ata}
A.~Allais and M.~Mezei, ``{Some results on the shape dependence of entanglement
  and Rényi entropies},''
  \href{http://dx.doi.org/10.1103/PhysRevD.91.046002}{{\em Phys. Rev.} {\bf
  D91} (2015) no.~4, 046002},
\href{http://arxiv.org/abs/1407.7249}{{\tt arXiv:1407.7249 [hep-th]}}.
%%CITATION = ARXIV:1407.7249;%%.

\bibitem{Lewkowycz:2014jia}
A.~Lewkowycz and E.~Perlmutter, ``{Universality in the geometric dependence of
  Renyi entropy},'' \href{http://dx.doi.org/10.1007/JHEP01(2015)080}{{\em JHEP}
  {\bf 01} (2015)  080},
\href{http://arxiv.org/abs/1407.8171}{{\tt arXiv:1407.8171 [hep-th]}}.
%%CITATION = ARXIV:1407.8171;%%.

\bibitem{Rosenhaus:2014zza}
V.~Rosenhaus and M.~Smolkin, ``{Entanglement Entropy for Relevant and Geometric
  Perturbations},'' \href{http://dx.doi.org/10.1007/JHEP02(2015)015}{{\em JHEP}
  {\bf 02} (2015)  015},
\href{http://arxiv.org/abs/1410.6530}{{\tt arXiv:1410.6530 [hep-th]}}.
%%CITATION = ARXIV:1410.6530;%%.

\bibitem{Mezei:2014zla}
M.~Mezei, ``{Entanglement entropy across a deformed sphere},''
  \href{http://dx.doi.org/10.1103/PhysRevD.91.045038}{{\em Phys. Rev.} {\bf
  D91} (2015) no.~4, 045038},
\href{http://arxiv.org/abs/1411.7011}{{\tt arXiv:1411.7011 [hep-th]}}.
%%CITATION = ARXIV:1411.7011;%%.

\bibitem{Carmi:2015dla}
D.~Carmi, ``{On the Shape Dependence of Entanglement Entropy},''
  \href{http://dx.doi.org/10.1007/JHEP12(2015)043}{{\em JHEP} {\bf 12} (2015)
  043},
\href{http://arxiv.org/abs/1506.07528}{{\tt arXiv:1506.07528 [hep-th]}}.
%%CITATION = ARXIV:1506.07528;%%.

\bibitem{Faulkner:2015csl}
T.~Faulkner, R.~G. Leigh, and O.~Parrikar, ``{Shape Dependence of Entanglement
  Entropy in Conformal Field Theories},''
  \href{http://dx.doi.org/10.1007/JHEP04(2016)088}{{\em JHEP} {\bf 04} (2016)
  088},
\href{http://arxiv.org/abs/1511.05179}{{\tt arXiv:1511.05179 [hep-th]}}.
%%CITATION = ARXIV:1511.05179;%%.

\bibitem{Leichenauer:2016rxw}
S.~Leichenauer, M.~Moosa, and M.~Smolkin, ``{Dynamics of the Area Law of
  Entanglement Entropy},''
  \href{http://dx.doi.org/10.1007/JHEP09(2016)035}{{\em JHEP} {\bf 09} (2016)
  035},
\href{http://arxiv.org/abs/1604.00388}{{\tt arXiv:1604.00388 [hep-th]}}.
%%CITATION = ARXIV:1604.00388;%%.

\bibitem{Dong-Renyi}
X.~Dong, ``{The Gravity Dual of Renyi Entropy},''
  \href{http://dx.doi.org/10.1038/ncomms12472}{{\em Nature Commun.} {\bf 7}
  (2016)  12472},
\href{http://arxiv.org/abs/1601.06788}{{\tt arXiv:1601.06788 [hep-th]}}.
%%CITATION = ARXIV:1601.06788;%%.

\bibitem{RT}
S.~Ryu and T.~Takayanagi, ``{Holographic derivation of entanglement entropy
  from AdS/CFT},'' \href{http://dx.doi.org/10.1103/PhysRevLett.96.181602}{{\em
  Phys. Rev. Lett.} {\bf 96} (2006)  181602},
\href{http://arxiv.org/abs/hep-th/0603001}{{\tt arXiv:hep-th/0603001
  [hep-th]}}.
%%CITATION = HEP-TH/0603001;%%.

\bibitem{hubeny2007covariant}
V.~E. Hubeny, M.~Rangamani, and T.~Takayanagi, ``A covariant holographic
  entanglement entropy proposal,'' {\em Journal of High Energy Physics} {\bf
  2007} (2007) no.~07, 062.

\bibitem{Lewkowycz:2013nqa}
A.~Lewkowycz and J.~Maldacena, ``{Generalized gravitational entropy},''
  \href{http://dx.doi.org/10.1007/JHEP08(2013)090}{{\em JHEP} {\bf 08} (2013)
  090},
\href{http://arxiv.org/abs/1304.4926}{{\tt arXiv:1304.4926 [hep-th]}}.
%%CITATION = ARXIV:1304.4926;%%.

\bibitem{Dong:2016hjy}
X.~Dong, A.~Lewkowycz, and M.~Rangamani, ``{Deriving covariant holographic
  entanglement},'' \href{http://dx.doi.org/10.1007/JHEP11(2016)028}{{\em JHEP}
  {\bf 11} (2016)  028},
\href{http://arxiv.org/abs/1607.07506}{{\tt arXiv:1607.07506 [hep-th]}}.
%%CITATION = ARXIV:1607.07506;%%.

\bibitem{Wilde-14}
M.~M. Wilde, A.~Winter, and D.~Yang, ``{Strong Converse for the Classical
  Capacity of Entanglement-Breaking and Hadamard Channels via a Sandwiched
  Renyi Relative Entropy},''
  \href{http://dx.doi.org/10.1007/s00220-014-2122-x}{{\em Commun. Math. Phys.}
  {\bf 331} (2014) no.~2, 593--622},
\href{http://arxiv.org/abs/1306.1586}{{\tt arXiv:1306.1586 [quant-ph]}}.
%%CITATION = ARXIV:1306.1586;%%.

\bibitem{sand13}
M.~{M{\"u}ller-Lennert}, F.~{Dupuis}, O.~{Szehr}, S.~{Fehr}, and
  M.~{Tomamichel}, \href{http://dx.doi.org/10.1063/1.4838856}{``{On quantum
  R{\'e}nyi entropies: A new generalization and some properties},''{\em Journal
  of Mathematical Physics} {\bf 54} (Dec, 2013)  122203--122203},
  \href{http://arxiv.org/abs/1306.3142}{{\tt arXiv:1306.3142 [quant-ph]}}.

\bibitem{FL-13}
R.~L. {Frank} and E.~H. {Lieb},
  \href{http://dx.doi.org/10.1063/1.4838835}{``{Monotonicity of a relative
  R{\'e}nyi entropy},''{\em Journal of Mathematical Physics} {\bf 54} (Dec,
  2013)  122201--122201}, \href{http://arxiv.org/abs/1306.5358}{{\tt
  arXiv:1306.5358 [math-ph]}}.

\bibitem{beigi}
S.~{Beigi}, \href{http://dx.doi.org/10.1063/1.4838855}{``{Sandwiched R{\'e}nyi
  divergence satisfies data processing inequality},''{\em Journal of
  Mathematical Physics} {\bf 54} (Dec, 2013)  122202--122202},
  \href{http://arxiv.org/abs/1306.5920}{{\tt arXiv:1306.5920 [quant-ph]}}.

\bibitem{MO-15}
M.~{Mosonyi} and T.~{Ogawa},
  \href{http://dx.doi.org/10.1007/s00220-014-2248-x}{``{Quantum Hypothesis
  Testing and the Operational Interpretation of the Quantum R{\'e}nyi Relative
  Entropies},''{\em Communications in Mathematical Physics} {\bf 334} (Mar,
  2015)  1617--1648}, \href{http://arxiv.org/abs/1309.3228}{{\tt
  arXiv:1309.3228 [quant-ph]}}.

\bibitem{araki}
H.~Araki, \href{http://dx.doi.org/10.1143/PTP.32.956}{``{Type of von Neumann
  Algebra Associated with Free Field},''{\em Progress of Theoretical Physics}
  {\bf 32} (12, 1964)  956--965},
  \href{http://arxiv.org/abs/http://oup.prod.sis.lan/ptp/article-pdf/32/6/956/5311286/32-6-956.pdf}{{\tt
  http://oup.prod.sis.lan/ptp/article-pdf/32/6/956/5311286/32-6-956.pdf}}.
  \url{https://doi.org/10.1143/PTP.32.956}.

\bibitem{Longo}
R.~Longo, ``{Algebraic and modular structure of von Neumann algebras of
  physics},''
{\em Commun. Math. Phys.} {\bf 38} (1982)  551.
%%CITATION = CMPHA,38,551;%%.

\bibitem{Fredenhagen1985}
K.~Fredenhagen, \href{http://dx.doi.org/10.1007/BF01206179}{``On the modular
  structure of local algebras of observables,''{\em Communications in
  Mathematical Physics} {\bf 97} (Mar, 1985)  79--89}.
  \url{https://doi.org/10.1007/BF01206179}.

\bibitem{zbMATH03511002}
H.~{Araki}, ``{Relative entropy of states of von Neumann algebras.},'' {\em
  {Publ. Res. Inst. Math. Sci.}} {\bf 11} (1976)  809--833.

\bibitem{RCP25_1975__22__A1_0}
H.~Araki, ``Inequalities in Von Neumann Algebras,'' {\em Les rencontres
  physiciens-math\'ematiciens de Strasbourg -RCP25} {\bf 22} (1975)  1--25.
  \url{http://www.numdam.org/item/RCP25_1975__22__A1_0}. talk:1.

\bibitem{Witten:2018lha}
E.~Witten, ``{APS Medal for Exceptional Achievement in Research: Invited
  article on entanglement properties of quantum field theory},''
  \href{http://dx.doi.org/10.1103/RevModPhys.90.045003}{{\em Rev. Mod. Phys.}
  {\bf 90} (2018) no.~4, 045003},
\href{http://arxiv.org/abs/1803.04993}{{\tt arXiv:1803.04993 [hep-th]}}.
%%CITATION = ARXIV:1803.04993;%%.

\bibitem{Lashkari-sand}
N.~Lashkari, ``{Constraining Quantum Fields using Modular Theory},''
  \href{http://dx.doi.org/10.1007/JHEP01(2019)059}{{\em JHEP} {\bf 01} (2019)
  059},
\href{http://arxiv.org/abs/1810.09306}{{\tt arXiv:1810.09306 [hep-th]}}.
%%CITATION = ARXIV:1810.09306;%%.

\bibitem{Casini:2011kv}
H.~Casini, M.~Huerta, and R.~C. Myers, ``{Towards a derivation of holographic
  entanglement entropy},''
  \href{http://dx.doi.org/10.1007/JHEP05(2011)036}{{\em JHEP} {\bf 05} (2011)
  036},
\href{http://arxiv.org/abs/1102.0440}{{\tt arXiv:1102.0440 [hep-th]}}.
%%CITATION = ARXIV:1102.0440;%%.

\bibitem{Ugajin:2018rwd}
T.~Ugajin, ``{Perturbative expansions of Rényi relative divergences and
  holography},''
\href{http://arxiv.org/abs/1812.01135}{{\tt arXiv:1812.01135 [hep-th]}}.
%%CITATION = ARXIV:1812.01135;%%.

\bibitem{add-ref-1}
A.~Bernamonti, F.~Galli, R.~C. Myers, and J.~Oppenheim, ``{Holographic second
  laws of black hole thermodynamics},''
  \href{http://dx.doi.org/10.1007/JHEP07(2018)111}{{\em JHEP} {\bf 07} (2018)
  111},
\href{http://arxiv.org/abs/1803.03633}{{\tt arXiv:1803.03633 [hep-th]}}.
%%CITATION = ARXIV:1803.03633;%%.

\bibitem{May:2018tir}
A.~May and E.~Hijano, ``{The holographic entropy zoo},''
  \href{http://dx.doi.org/10.1007/JHEP10(2018)036}{{\em JHEP} {\bf 10} (2018)
  036},
\href{http://arxiv.org/abs/1806.06077}{{\tt arXiv:1806.06077 [hep-th]}}.
%%CITATION = ARXIV:1806.06077;%%.

\bibitem{W-noether-entropy}
R.~M. Wald, ``{Black hole entropy is the Noether charge},''
  \href{http://dx.doi.org/10.1103/PhysRevD.48.R3427}{{\em Phys. Rev.} {\bf D48}
  (1993) no.~8, R3427--R3431},
\href{http://arxiv.org/abs/gr-qc/9307038}{{\tt arXiv:gr-qc/9307038 [gr-qc]}}.
%%CITATION = GR-QC/9307038;%%.

\bibitem{IW-noether-entropy}
V.~Iyer and R.~M. Wald, ``{Some properties of Noether charge and a proposal for
  dynamical black hole entropy},''
  \href{http://dx.doi.org/10.1103/PhysRevD.50.846}{{\em Phys. Rev.} {\bf D50}
  (1994)  846--864},
\href{http://arxiv.org/abs/gr-qc/9403028}{{\tt arXiv:gr-qc/9403028 [gr-qc]}}.
%%CITATION = GR-QC/9403028;%%.

\bibitem{LW}
J.~Lee and R.~M. Wald, ``{Local symmetries and constraints},''
\href{http://dx.doi.org/10.1063/1.528801}{{\em J. Math. Phys.} {\bf 31} (1990)
  725--743}.
%%CITATION = JMAPA,31,725;%%.

\bibitem{WZ}
R.~M. Wald and A.~Zoupas, ``{A General definition of 'conserved quantities' in
  general relativity and other theories of gravity},''
  \href{http://dx.doi.org/10.1103/PhysRevD.61.084027}{{\em Phys. Rev.} {\bf
  D61} (2000)  084027},
\href{http://arxiv.org/abs/gr-qc/9911095}{{\tt arXiv:gr-qc/9911095 [gr-qc]}}.
%%CITATION = GR-QC/9911095;%%.

\bibitem{HW}
S.~Hollands and R.~M. Wald, ``{Stability of Black Holes and Black Branes},''
  \href{http://dx.doi.org/10.1007/s00220-012-1638-1}{{\em Commun. Math. Phys.}
  {\bf 321} (2013)  629--680},
\href{http://arxiv.org/abs/1201.0463}{{\tt arXiv:1201.0463 [gr-qc]}}.
%%CITATION = ARXIV:1201.0463;%%.

\bibitem{Lindblad1974}
G.~Lindblad, ``Expectations and entropy inequalities for finite quantum
  systems,''{\em Communications in Mathematical Physics} {\bf 39} (Jun, 1974)
  111--119.

\bibitem{k}
R.~Kubo, ``Statistical-Mechanical Theory of Irreversible Processes. I. General
  Theory and Simple Applications to Magnetic and Conduction Problems,''
  \href{http://dx.doi.org/10.1143/JPSJ.12.570}{{\em Journal of the Physical
  Society of Japan} {\bf 12} (1957) no.~6, 570--586},
  \href{http://arxiv.org/abs/https://doi.org/10.1143/JPSJ.12.570}{{\tt
  https://doi.org/10.1143/JPSJ.12.570}}.
  \url{https://doi.org/10.1143/JPSJ.12.570}.

\bibitem{ms}
P.~C. Martin and J.~Schwinger,
  \href{http://dx.doi.org/10.1103/PhysRev.115.1342}{``Theory of Many-Particle
  Systems. I,''{\em Phys. Rev.} {\bf 115} (Sep, 1959)  1342--1373}.
  \url{https://link.aps.org/doi/10.1103/PhysRev.115.1342}.

\bibitem{Haag1967}
R.~Haag, N.~M. Hugenholtz, and M.~Winnink, ``On the equilibrium states in
  quantum statistical mechanics,''{\em Communications in Mathematical Physics}
  {\bf 5} (Jun, 1967)  215--236.

\bibitem{Witten:1998qj}
E.~Witten, ``{Anti-de Sitter space and holography},''
  \href{http://dx.doi.org/10.4310/ATMP.1998.v2.n2.a2}{{\em Adv. Theor. Math.
  Phys.} {\bf 2} (1998)  253},
\href{http://arxiv.org/abs/hep-th/9802150}{{\tt arXiv:hep-th/9802150
  [hep-th]}}.
%%CITATION = HEP-TH/9802150;%%.

\bibitem{Marolf:2017kvq}
D.~Marolf, O.~Parrikar, C.~Rabideau, A.~Izadi~Rad, and M.~Van~Raamsdonk,
  ``{From Euclidean Sources to Lorentzian Spacetimes in Holographic Conformal
  Field Theories},'' \href{http://dx.doi.org/10.1007/JHEP06(2018)077}{{\em
  JHEP} {\bf 06} (2018)  077},
\href{http://arxiv.org/abs/1709.10101}{{\tt arXiv:1709.10101 [hep-th]}}.
%%CITATION = ARXIV:1709.10101;%%.

\bibitem{Gubser:1998bc}
S.~S. Gubser, I.~R. Klebanov, and A.~M. Polyakov, ``{Gauge theory correlators
  from noncritical string theory},''
  \href{http://dx.doi.org/10.1016/S0370-2693(98)00377-3}{{\em Phys. Lett.} {\bf
  B428} (1998)  105},
\href{http://arxiv.org/abs/hep-th/9802109}{{\tt arXiv:hep-th/9802109
  [hep-th]}}.
%%CITATION = HEP-TH/9802109;%%.

\bibitem{Akers:2018fow}
C.~Akers and P.~Rath, ``{Holographic Renyi Entropy from Quantum Error
  Correction},''
\href{http://arxiv.org/abs/1811.05171}{{\tt arXiv:1811.05171 [hep-th]}}.
%%CITATION = ARXIV:1811.05171;%%.

\bibitem{Dong:2018seb}
X.~Dong, D.~Harlow, and D.~Marolf, ``{Flat entanglement spectra in fixed-area
  states of quantum gravity},''
\href{http://arxiv.org/abs/1811.05382}{{\tt arXiv:1811.05382 [hep-th]}}.
%%CITATION = ARXIV:1811.05382;%%.

\bibitem{1982NuPhB}
L.~F. {Abbott} and S.~{Deser},
  \href{http://dx.doi.org/10.1016/0550-3213(82)90049-9}{``{Stability of gravity
  with a cosmological constant},''{\em Nuclear Physics B} {\bf 195} (Feb.,
  1982)  76--96}.

\bibitem{Woolgar:1994ar}
E.~Woolgar, ``{The Positivity of energy for asymptotically anti-de Sitter
  space-times},'' \href{http://dx.doi.org/10.1088/0264-9381/11/7/022}{{\em
  Class. Quant. Grav.} {\bf 11} (1994)  1881--1900},
\href{http://arxiv.org/abs/gr-qc/9404019}{{\tt arXiv:gr-qc/9404019 [gr-qc]}}.
%%CITATION = GR-QC/9404019;%%.

\bibitem{Bousso:2015mna}
R.~Bousso, Z.~Fisher, S.~Leichenauer, and A.~C. Wall, ``{Quantum focusing
  conjecture},'' \href{http://dx.doi.org/10.1103/PhysRevD.93.064044}{{\em Phys.
  Rev.} {\bf D93} (2016) no.~6, 064044},
\href{http://arxiv.org/abs/1506.02669}{{\tt arXiv:1506.02669 [hep-th]}}.
%%CITATION = ARXIV:1506.02669;%%.

\bibitem{Casini:2017roe}
H.~Casini, E.~Teste, and G.~Torroba, ``{Modular Hamiltonians on the null plane
  and the Markov property of the vacuum state},''
  \href{http://dx.doi.org/10.1088/1751-8121/aa7eaa}{{\em J. Phys.} {\bf A50}
  (2017) no.~36, 364001},
\href{http://arxiv.org/abs/1703.10656}{{\tt arXiv:1703.10656 [hep-th]}}.
%%CITATION = ARXIV:1703.10656;%%.

\bibitem{add-ref-2}
H.~Casini, R.~Medina, I.~Salazar~Landea, and G.~Torroba, ``{Renyi relative
  entropies and renormalization group flows},''
  \href{http://dx.doi.org/10.1007/JHEP09(2018)166}{{\em JHEP} {\bf 09} (2018)
  166},
\href{http://arxiv.org/abs/1807.03305}{{\tt arXiv:1807.03305 [hep-th]}}.
%%CITATION = ARXIV:1807.03305;%%.

\bibitem{hayden2013holographic}
P.~Hayden, M.~Headrick, and A.~Maloney, ``Holographic mutual information is
  monogamous,'' {\em Physical Review D} {\bf 87} (2013) no.~4, 046003.

\bibitem{bao2015holographic}
N.~Bao, S.~Nezami, H.~Ooguri, B.~Stoica, J.~Sully, and M.~Walter, ``The
  holographic entropy cone,'' {\em Journal of High Energy Physics} {\bf 2015}
  (2015) no.~9, 130.

\end{thebibliography}\endgroup
\end{document}